\documentclass[twocolumn]{aastex631}

\usepackage{subfigure}
\usepackage{graphicx}
\usepackage{rotating} 
\usepackage[autostyle=false, style=english]{csquotes}
\usepackage{natbib}
\usepackage{multirow}
\usepackage{appendix}
\begin{document}

\title{Unveiling the Dual Nature of V1180 Cas: UXor-like Dips and EXor-like Bursts Across a Decade}
\author[0009-0008-8490-8601]{Tarak Chand}
\affiliation{Aryabhatta Research Institute of Observational Sciences (ARIES), Manora Peak, Nainital-263001, India}
\affiliation{M.J.P. Rohilkhand University, Bareilly-243006, India}
\correspondingauthor{Tarak Chand}
\email{tarakchands@gmail.com, tarakchand@aries.res.in}

\author[0000-0001-5731-3057]{Saurabh Sharma}
\affiliation{Aryabhatta Research Institute of Observational Sciences (ARIES), Manora Peak, Nainital-263001, India}

\author[0000-0002-7434-9681]{Koshvendra Singh}
\affiliation{Department of Astronomy and Astrophysics, Tata Institute of Fundamental Research (TIFR), Mumbai-400005, India}

\author[0000-0001-8720-5612]{Joe P. Ninan}
\affiliation{Department of Astronomy and Astrophysics, Tata Institute of Fundamental Research (TIFR), Mumbai-400005, India}

\author[0000-0001-7650-1870]{Arpan Ghosh}
\affiliation{Instituto de Radioastronom{\'i}a y Astrof{\'i}sica, Universidad Nacional Aut{\'o}noma de M{\'e}xico, Antigua Carretera a P{\'a}tzcuaro 8701, Ex-Hda. San Jos{\'e} de la Huerta, Morelia, Michoac{\'a}n, M{\'e}xico C.P. 58089 }

\author[0000-0001-9312-3816]{Devendra K. Ojha}
\affiliation{Department of Astronomy and Astrophysics, Tata Institute of Fundamental Research (TIFR), Mumbai-400005, India}

\author[0000-0003-0295-6586]{Tapas Baug}
\affiliation{Satyendra Nath Bose National Centre for Basic Sciences (SNBNCBS), Block-JD, Sector-III, Salt Lake, Kolkata-700106, India}

\author[0000-0002-6688-0800]{D. K. Sahu}
\affiliation{Indian Institute of Astrophysics (IIA), II Block, Koramangala, Bangalore-560034, India}

\author{Bhuwan C. Bhatt}
\affiliation{Indian Institute of Astrophysics (IIA), II Block, Koramangala, Bangalore-560034, India}

\author{Pramod Kumar}
\affiliation{Indian Institute of Astrophysics (IIA), II Block, Koramangala, Bangalore-560034, India}

\author[0000-0002-6740-7425]{Ram K. Yadav}
\affiliation{National Astronomical Research Institute of Thailand (Public Organization), 260 Moo4, T. Donkaew, A. Maerim, Chiangmai-50180, Thailand}

\author[0000-0002-0151-2361]{Neelam Panwar }
\affiliation{Aryabhatta Research Institute of Observational Sciences (ARIES), Manora Peak, Nainital-263001, India}

\author[0000-0002-6586-936X]{Aayushi Verma}
\affiliation{Aryabhatta Research Institute of Observational Sciences (ARIES), Manora Peak, Nainital-263001, India}
\affiliation{M.J.P. Rohilkhand University, Bareilly-243006, India}

\author[0000-0002-0444-0439]{Harmeen Kaur}
\affiliation{Aryabhatta Research Institute of Observational Sciences (ARIES), Manora Peak, Nainital-263001, India}

\author[0009-0001-4144-2281]{Mamta}
\affiliation{Aryabhatta Research Institute of Observational Sciences (ARIES), Manora Peak, Nainital-263001, India}
\affiliation{Indian Institute of Technology (IIT), Roorkee, Roorkee-247667, India}

\author[0009-0009-6420-8058]{Manojit Chakraborty}
\affiliation{Aryabhatta Research Institute of Observational Sciences (ARIES), Manora Peak, Nainital-263001, India}
\affiliation{Indian Institute of Technology (IIT), Roorkee, Roorkee-247667, India}

\author[0009-0004-4776-6780]{Kartik Gokhe}
\affiliation{Aryabhatta Research Institute of Observational Sciences (ARIES), Manora Peak, Nainital-263001, India}

\author{Ajay Kumar Singh}
\affiliation{Department of Applied Physics/Physics, Bareilly College, Bareilly-243001, India}

\begin{abstract}

We present a detailed analysis of the long-term photometric and spectroscopic evolution of V1180 Cas over a decade, aiming to identify the dominant mechanisms behind its variability. We combine multi-band light curves from 1999 to 2025 with over 30 epochs of optical to near-infrared spectroscopy (0.5-2.5 $\mu$m), analyzing variability patterns, color behavior, and emission line diagnostics. We investigate the temporal evolution of accretion and outflow indicators and their correlation with photometric states. The light curve reveals a transition from sporadic early dimming events to a quasi-periodic pattern since 2018, with eleven major dips showing asymmetry and stochastic sub-structure. Color-magnitude diagrams show classic UXor-like blueing during deep minima, while near-infrared and mid-infrared color changes indicate thermal evolution of disk. Spectroscopic analysis reveals persistent hydrogen, Ca II, He I, and forbidden line emission. Accretion diagnostics track photometric variability, and forbidden lines often intensify during dips, implying a physical link between extinction and outflows. Estimated accretion rates range from $\sim10^{-8}–10^{-7}$ $M_\odot$yr$^{-1}$; the outflow rate and density diagnostics are consistent with disk winds and shock-excited jets. V1180 Cas demonstrates dual-mode variability driven by both variable circumstellar extinction and episodic accretion events. The hybrid UXor/EXor behavior, combined with evolving disk signatures and persistent outflows, suggests a young stellar object undergoing coupled accretion–extinction–outflow evolution. Continued monitoring will be essential to fully resolve the physical processes shaping its variability.
\end{abstract}


\section{Introduction} \label{sec:intro}

Young stellar objects (YSOs) often exhibit variability in brightness due to processes related to accretion and circumstellar material. A distinct subclass of these objects, known as young eruptive stars, is characterized by large-amplitude variability, ranging from 2 to 5 magnitudes (mag), observed at optical and infrared (IR) wavelengths over timescales spanning months to centuries \citep{Hartmann_1996ARA&A..34..207H, Audard_2014prpl.conf..387A}. These stars undergo episodic outbursts during which their luminosity can increase by up to two orders of magnitude. Such outbursts are attributed to a sudden increase in mass accretion rates from the circumstellar disk onto the stellar surface, rising from quiescent levels of $\sim$10$^{-10}$-10$^{-9}$ $M_{\odot}$yr$^{-1}$ to as high as 10$^{-5}$--10$^{-4}$ $M_{\odot}$yr$^{-1}$ during outburst events. Growing evidence suggests that this phenomenon of episodic accretion is a fundamental aspect of star formation \citep[][and references therein]{Fischer_2023ASPC..534..355F}.

Young eruptive stars are traditionally classified into two main types based on their photometric and spectroscopic outburst characteristics: FU Orionis-type (FUors) and EX Lupi-type (EXors) \citep{Herbig_1977ApJ...217..693H, Herbig_2008AJ....135..637H}. FUors exhibit large-amplitude (up to 5 mag) outbursts that can persist for decades to centuries and may not repeat. In contrast, EXors undergo lower-energy, recurrent outbursts lasting from months to about a year. Spectroscopically, FUors are dominated by absorption features during outburst, including Ca II IR triplet (8498 \r{A}, 8542 \r{A} and 8662 \r{A}), Na I D(5900 \r{A}), and CO bandhead absorption at 2.2935 and 2.3227 $\mu$m. EXors, on the other hand, display spectra rich in emission lines at maximum light, notably Na I (2.2060 $\mu$m) and CO bandhead emission, which originate primarily from the hot inner disk but can also arise from the outer, cooler regions of the accretion columns \citep{2018ApJ...861..145C}.

Other classes of YSOs that exhibit large-amplitude photometric variability comparable to FUors and EXors include the UXors \citep{Herbst_1994AJ....108.1906H} and so-called \enquote{dippers} \citep{1999A&A...349..619B, Ansdell_2016ApJ...816...69A, Kennedy_2017RSOS....460652K}. Their variability arises primarily from irregular brightness dips caused by circumstellar dust extinction \citep{Herbst_1994AJ....108.1906H, Natta_2000A&A...364..633N, Dullemond_2003ApJ...594L..47D, Kennedy_2017RSOS....460652K}. UXors are intermediate- (Herbig Ae/Be) and low-mass (T Tauri) stars, in which hydrodynamic turbulence can lift clumps of gas and dust above the disk, producing transient extinction events \citep{Abraham_2018ApJ...853...28A, 2025A&A...694A.257T}. A hallmark of UXor-type fading is the \enquote{blueing effect} during the minima of the brightness. In contrast, dippers are typically low-mass stars whose light curves (LCs) display quasi-periodic or aperiodic flux drops, attributed to short-timescale (few days) occultations by dusty inner disk structures, accretion streams, or magnetically induced disk warps \citep{2021A&A...651A..44R, 2023MNRAS.521.1700M}. Unlike UXors, dippers generally do not show \enquote{blueing effect} during the dip in the LC \citep{2022AJ....163..263H}.
Prototypes of these classes include UX Ori and AA Tau, with other well-studied cases such as RR Tau, GM Cep, and VV Ser \citep{Bouvier_2003A&A...409..169B, Bedell_2011AJ....142..164B, Semkov_2015PASA...32...11S, Garcia_2016MNRAS.456..156G}. Some young stars, such as V2492 Cyg and V582 Aur, appear to show hybrid variability, with contributions from both accretion-driven outbursts and variable line-of-sight extinction \citep{Hillenbrand_2013AJ....145...59H, Abraham_2018ApJ...853...28A, 2019ApJ...873..130Z}.

\begin{figure*}
    \includegraphics[width=\textwidth]{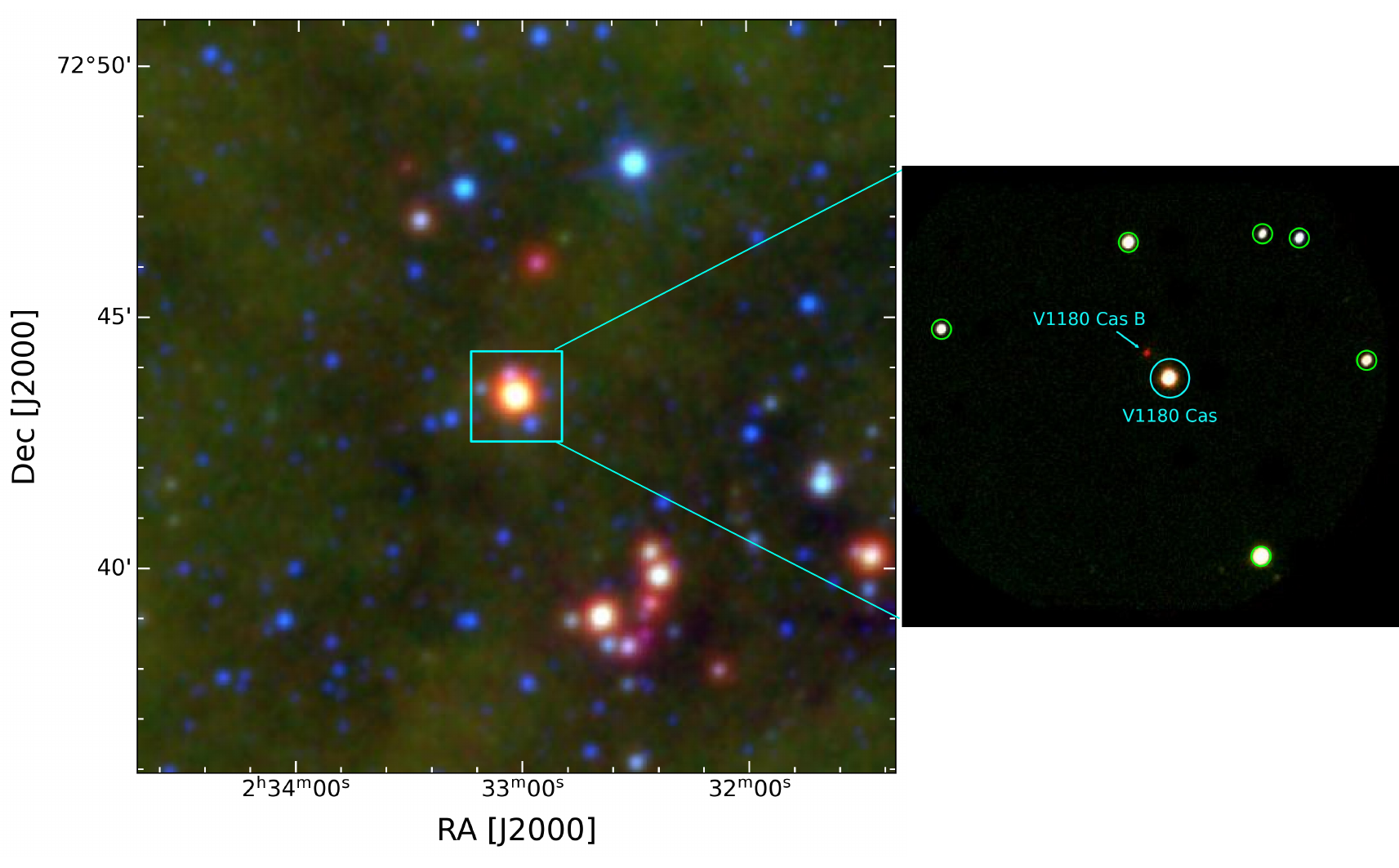}
    \caption{Left panel: Color-composite image obtained using the $W4$ (22 $\mu$m), $W3$ (12 $\mu$m), and $W2$ (4.6 $\mu$m) WISE images, shown as red, green, and blue colors, respectively. The Field of View (FoV) of this color-composite image is $\sim$10$^{\prime}$ $\times$ 10$^{\prime}$ around the V1180 Cas. The cyan color box shows the FoV of TANSPEC. Right panel: The color-composite image comprises TANSPEC $K$, $H$, and $J$ band images shown as red, green, and blue, respectively. The cyan circle shows the location of the V1180 Cas, while reference stars, which are used to calculate offsets to calibrate the instrumental magnitudes (see Section \ref{subsubsec:observed}), are marked with green circles.}
    \label{fig:rgb}
\end{figure*}
V1180 Cas is a similar object exhibiting multiple, large, and long-duration photometric dips. V1180 Cas was first identified as a YSO by \citet{Kun_1994A&A...292..249K} during an H${\alpha}$ survey of YSOs in the Lynds 1340 molecular cloud. Using a decade-long (1999 to 2011) photometric dataset, \cite{Kun_2011} reported that V1180 Cas exhibits $\sim$6 mag of variation in its $I_{c}$ LC. The authors found that the star becomes bluer in a dim state, attributing this to an increase in the fraction of scattered light in the optical fluxes. 

Through $K_s$-band images, \cite{Kun_2011} also identified an optically invisible companion star, V1180 Cas B, with a projected separation of 3600 au. The authors estimated an optical extinction $A_V = 4.3$ mag, using the ($I_{c}$ - $J$) color index and assuming the spectral type of the star to be K7 ($M_{*} = 0.8$ $M_{\odot}$, $R_{*} = 2$ $R_{\odot}$). From the luminosity of the Ca II ($8542$ \r{A}) emission, they calculated an accretion rate $\dot{M}_{acc} > 1.6 \times 10^{-7}$ $M_{\odot}$yr$^{-1}$ in the bright state ($I$ $\sim$15 mag), which decreased by approximately a factor of ten when its $I_{c}$-band brightness got dimmer by $\approx$3 mag in year 2005.

Later, \cite{Antoniucci_2014A&A...565L...7A} calculated accretion rates using more than ten emission spectral lines, finding a median accretion rate of $\sim$$3 \times 10^{-8}$ $M_{\odot}$yr$^{-1}$ in bright state ($I$ $\sim$15.5 mag).
They also calculated the mass-loss rate to be $\sim$$4 \times 10^{-9}$ $M_{\odot}$yr$^{-1}$ and $\sim$$4 \times 10^{-10}$ $M_{\odot}$yr$^{-1}$, using atomic forbidden lines and H$_2$ line at 21218 \r{A}, respectively.
Utilizing $H_2$ and $K$-band images, they observed a jet (a series of knots) in the region of V1180 Cas, though they were uncertain of the jet's driving source due to insufficient spatial resolution.

Recently, \cite{Mutafov_2022RAA....22l5014M} reported several dips in the LC of V1180 Cas using long-term (September 2011 to February 2022) $V$, $R$, and $I$ band observations. The largest dip observed was $\Delta$$I$ $\approx$5 mag. They also noted the \enquote{blueing effect} during the minima and concluded that the variability of V1180 Cas was a combination of highly variable accretion and occultation by the circumstellar clumps.

In this paper, we present the results of our long-term optical and near-infrared (NIR) spectroscopic and photometric monitoring of V1180 Cas, conducted as part of our ongoing program, Monitoring FU Orionis and EX Orionis Sources (MFES; \citealt{Ninan_2015ApJ...815....4N, Ninan_2016ApJ...825...65N, Ghosh_2022ApJ...926...68G, Ghosh_2023ApJ...954...82G, Ghosh_2023JApA...44...50G, Singh_2023JApA...44...58S, Singh_2024ApJ...968...88S}). By supplementing these observations with archival photometric data from various catalogs, our goal is to investigate the origin of the pronounced brightness variations exhibited by V1180 Cas.
The Figure \ref{fig:rgb} shows the color-composite image of V1180 Cas. The left panel is generated using mid-infrared (MIR) $W4$ (22 $\mu$m), $W3$ (12 $\mu$m), and $W2$ (4.6 $\mu$m) WISE images, as red, green, and blue colors, respectively. The right panel shows a color-composite image from the current TIFR-ARIES Near Infrared Spectrometer (TANSPEC) observations, combining $K$, $H$, and $J$ bands as red, green, and blue, respectively. The cyan circle marks the location of V1180 Cas, while reference stars used to compute magnitude calibration offsets (see Section \ref{sec:obs}) are indicated with green circles. The close companion, V1180 Cas B---invisible in the optical bands---is also marked in the image. The extended yellow-red emission surrounding V1180 Cas in the WISE image suggests the presence of significant MIR emission ($W4$ and $W3$), consistent with a dense circumstellar environment composed of gas and dust, likely in the form of a disk, envelope, and/or outflowing material such as winds or jets.

The paper is structured as follows: Section \ref{sec:obs} details the observations and the methods used for photometric and spectroscopic data reduction. In Section \ref{sec:result}, we outline the key findings derived from our analysis of the data. Section \ref{sec:discussion} explores the underlying causes of the observed variability and the possible nature of V1180 Cas. Finally, we summarize and conclude our major findings from this study in Section \ref{sec:summ}.

\begin{table*}
   \centering
    \caption{Log of Photometric and Spectroscopic Observations.}
    \begin{tabular}{c c c c c}
        \hline
        \hline
        Date &       Data Type &    Grisms/Filters &  Exp$\times$Frames & Instrument \\
        (YYYYMMDD) & & & &\\
        
        \hline
20150818&   Spectra/Image &        G8/V, R, I &             1800/250$\times$3, 120$\times$3, 30$\times$3 &      HFOSC \\
20150927&         Spectra &                G7 &                                2800 &      HFOSC \\
20151014&   Spectra/Image &        G8/V, R, I &               1800/90$\times$3, 40$\times$3, 20$\times$3 &      HFOSC \\
20151016&           Image &           V, R, I &                 300$\times$3, 180$\times$3, 180$\times$3 &   DFOT 512 \\
20151102&           Image &           V, R, I &                 300$\times$3, 180$\times$3, 180$\times$3 &   DFOT 512 \\
20151110&   Spectra/Image &        G8/V, R, I &              2400/120$\times$3, 60$\times$3, 60$\times$3 &      HFOSC \\
20151118&           Image &           V, R, I &                 300$\times$3, 180$\times$3, 180$\times$3 &   DFOT 512 \\
20160102&           Image &           V, R, I &                 300$\times$3, 180$\times$3, 180$\times$3 &   DFOT 512 \\ 
20160118&           Image &           V, R, I &                 300$\times$3, 180$\times$3, 180$\times$3 &   DFOT 512 \\ 
20160120&         Spectra &                G8 &                                2700 &      HFOSC \\
20160203&    Spectra/Image &        G8/V, R, I &              1800/120$\times$3, 20$\times$3, 20$\times$3 &      HFOSC \\
20160929&           Image &           V, R, I &                 300$\times$3, 180$\times$3, 180$\times$3 &    DFOT 2K \\
20161006&         Spectra &            G7 &                          2400 &      HFOSC \\
20161008&           Image &           V, R, I &                 300$\times$3, 180$\times$3, 180$\times$3 &   DFOT 512 \\
20161115&           Image &           V, R, I &                 300$\times$3, 180$\times$3, 180$\times$3 &   DFOT 512 \\ 
20161208&   Spectra/Image &    G7, G8/V, R, I &            2400, 2400/180, 100, 100 &      HFOSC \\
20161210&   Spectra/Image &        G7/V, R, I &            1800/300$\times$3, 180$\times$3, 180$\times$3 &      HFOSC/DFOT 512 \\ 
20170102&         Spectra &            G7, G8 &                          2400, 2400 &      HFOSC \\
20170918&   Spectra/Image &    G7, G8/V, R, I &         1800, 1800/40$\times$4, 30$\times$3, 30$\times$3 &      HFOSC \\
20170815&           Image &           V, R, I &                 300$\times$3, 180$\times$3, 180$\times$3 &    DFOT 2K \\
20171024&   Spectra/Image &    G7, G8/V, R, I &         2400, 1800/50$\times$2, 30$\times$3, 30$\times$3 &      HFOSC \\
20171111&           Image &           V, R, I &                 300$\times$3, 180$\times$3, 180$\times$3 &    DFOT 2K \\
20171126&         Spectra &            G7, G8 &                         1800, 3000 &      HFOSC \\
20171210&           Image &           V, R, I &                 300$\times$3, 180$\times$3, 180$\times$3 &    DFOT 2K \\ 
20180218&         Spectra &            G7, G8 &                          3600, 3600 &      HFOSC \\
20181001&           Image &           V, R, I &                 300$\times$3, 120$\times$3, 120$\times$3 &    DFOT 2K \\ 
20181006&           Image &           V, R, I &                 300$\times$3, 180$\times$3, 180$\times$3 &    DFOT 2K \\ 
20181121&   Spectra/Image &    G7, G8/V, R, I &         2400, 1800/40$\times$3, 20$\times$3, 20$\times$3 &      HFOSC \\
20181122&           Image &           V, R, I &                 300$\times$3, 240$\times$3, 120$\times$3 &    DFOT 2K \\ 
20181212&         Spectra &            G7, G8 &                          2700, 2700 &      HFOSC \\
20181219&           Image &           V, R, I &                   300$\times$3, 60$\times$3, 60$\times$3 &   DFOT 512 \\
20181220&           Image &           V, R, I &                  300$\times$3, 150$\times$3, 90$\times$3 &   DFOT 512 \\ 
20181225&   Spectar/Image &    G7, G8/V, R, I &         2400, 1800/40$\times$3, 20$\times$3, 20$\times$3 &      HFOSC \\
20190619&           Image &           V, R, I &                    40$\times$3, 20$\times$4, 20$\times$2 &      HFOSC \\
20190927&         Spectra &                G8 &                                1800 &      HFOSC \\
20191008&         Spectra &            G7, G8 &                          3000, 2700 &      HFOSC \\
20191028&   Spectra/Image &        G7/V, R, I &            3600/250$\times$3, 200$\times$3, 200$\times$3 &      HFOSC \\
20200119&           Image &           V, R, I &                 300$\times$3, 240$\times$3, 240$\times$3 &      HFOSC \\
20200914&   Spectra/Image &    G7, G8/V, R, I &        3000, 3000/240$\times$2, 60$\times$2, 60$\times$2 &      HFOSC \\
20201030&         Spectra &                xd &                              180$\times$10 &    TANSPEC \\  
20210306&         Spectra &            G7, G8 &                          2100, 2100 &      HFOSC \\
20220207&         Spectra &                G8 &                                2700 &      HFOSC \\
20221102&   Spectra/Image &        xd/J, H, K &              180$\times$7/20$\times$3, 20$\times$3, 20$\times$3 &    TANSPEC \\ 
20240114&   Spectra/Image &  xd/J, H, K &  150$\times$9/20$\times$3, 20$\times$3, 20$\times$3 &    TANSPEC \\  
20240121&   Spectra/Image &  xd/J, H, K & 150$\times$10/10$\times$3, 10$\times$3, 10$\times$3 &    TANSPEC \\ 
20240122&   Spectra &        xd &             150$\times$10 &    TANSPEC \\ 
20240123&   Spectra/Image &  xd/J, H, K & 150$\times$10/10$\times$3, 10$\times$3, 10$\times$3 &    TANSPEC \\ 
20240124&   Spectra/Image &  xd/J, H, K & 150$\times$11/10$\times$3, 10$\times$3, 10$\times$3 &    TANSPEC \\ 
20240209&         Spectra &                xd &                              180$\times$10 &    TANSPEC \\ 
        \hline
    \end{tabular}
    \label{tab:obs_log}
\end{table*}

\begin{table*}[]
\centering
\caption{Archival photometric data used in this study.}
\begin{tabular}{ccc}
\hline
Survey & Filters & Reference \\
\hline
Gaia Data Release 3 (Gaia DR3\footnote{doi:10.26093/cds/vizier.1355}) &$G$, $G_{BP}$, $G_{RP}$ & \cite{Gaia_2016,2023gaia} \\
Zwicky Transient Facility (ZTF\footnote{doi:10.26131/IRSA598}) & $zg$, $zr$ & \cite{Bellm_2019PASP..131a8002B} \\
Asteroid Terrestrial-impact Last Alert System (ATLAS) & $c$, $o$ & \cite{Tonry_2018PASP..130f4505T} \\
Near-Earth Object Wide-field Infrared Survey Explorer (NEOWISE\footnote{doi:10.26131/IRSA144}) & $W1$ , $W2$  & \cite{Mainzer_2014ApJ...792...30M} \\
Two Micron All Sky Survey (2MASS) & $J$, $H$, $Ks$ & \cite{Skrutskie_2006AJ....131.1163S} \\
\hline
\end{tabular}
\label{tab:archive_data}
\end{table*}

\section{Observations and Data Reduction} \label{sec:obs}
\subsection{Photometric data}\label{subsec:Photo}
\subsubsection{Observed photometric data}\label{subsubsec:observed}
Decade-long (2015-2025) multi-epoch photometric observations of V1180 Cas in the optical $V$, $R$, and $I$ bands were carried out using the Himalayan Faint Object Spectrograph Camera (HFOSC), mounted on the 2 m Himalayan Chandra Telescope (HCT) and the 1.3 m Devasthal Fast Optical Telescope (DFOT). 
The HFOSC, with its 2k $\times$ 2k pixel CCD (2k CCD) camera, provides a FoV of $\sim$10$^{\prime}$ $\times$ 10$^{\prime}$ and a plate scale of $\sim$0.293$^{\prime \prime}$ per pixel (see HCT website\footnote{\url{https://www.iiap.res.in/centers/iao/facilities/hct/}}). 
The DFOT is equipped with a 2k CCD camera and a 512 $\times$ 512 pixel CCD (512 CCD) camera. The FoV for the 2k CCD is $\sim$18$^{\prime}$ $\times$ 18$^{\prime}$ with a plate scale of 0.541$^{\prime \prime}$ per pixel, while for the 512 CCD, the FoV is $\sim$5.5$^{\prime}$ $\times$ 5.5$^{\prime}$ with a plate scale of 0.641$^{\prime \prime}$ per pixel (for more details, see \citealt{Sagar_2012ASInC...4..173S}). 
Additionally, we acquired photometric data in the NIR bands ($J$, $H$, and $K$), using the H1RG array (1k $\times$ 1k pixels) of the TANSPEC mounted on the 3.6 m Devasthal Optical Telescope (DOT). TANSPEC, with H1RG array camera and 18 $\mu$m pixel size, has a FoV of $\sim$1$^{\prime}$ $\times$ 1$^{\prime}$ (for more details, see \citealt{Sharma_2022PASP..134h5002S}). The detailed observation log is provided in Table \ref{tab:obs_log}.

The data reduction (image cleaning, photometry, and astrometry) was performed using the standard procedure explained in \citet{2017MNRAS.467.2943S, 2020MNRAS.498.2309S} and \citet{Kaur_2020ApJ...896...29K}. 
The NIR instrumental magnitudes---obtained using TANSPEC---were calibrated using the color terms (slopes) from the calibration equations provided by \citet{Sharma_2022PASP..134h5002S}, along with zero-point offsets derived from 2MASS magnitudes of a few bright, isolated stars (marked in green circles in the right panel of Figure \ref{fig:rgb}).
The optical instrumental magnitudes--obtained using HFOSC---were calibrated using the photometric equations described in \citet{Kaur_2020ApJ...896...29K}, with the slope and zero-point offsets determined from the known magnitudes of nearby isolated bright stars.

Thereafter, the differential photometric technique was employed to generate the LCs (for details, see Section 3.4 of \citealt{Sinha_2020MNRAS.493..267S}). The standard deviations in the difference magnitudes for the selected pair of comparison stars are $\sim$0.05, 0.04, and 0.02 mag for the $V, R, $ and $I$ bands, respectively.

\subsubsection{Archival photometric data}\label{subsubsec:archive}

We also retrieved time-series optical and MIR photometric data from various archival sources for our analysis. The details are provided in Table \ref{tab:archive_data}.

\begin{figure*}
    \centering
    \subfigure{\includegraphics[width=.92\linewidth]{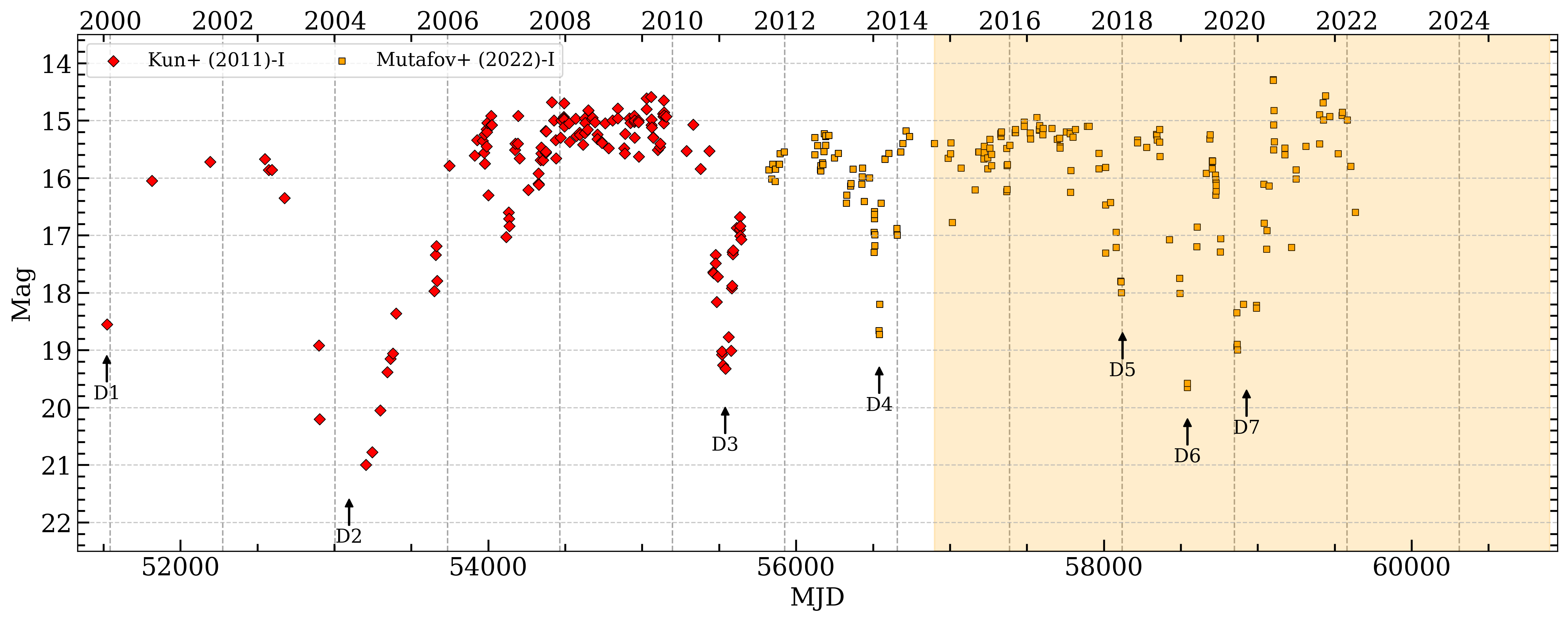}}
    \subfigure{\includegraphics[width=.95\linewidth]{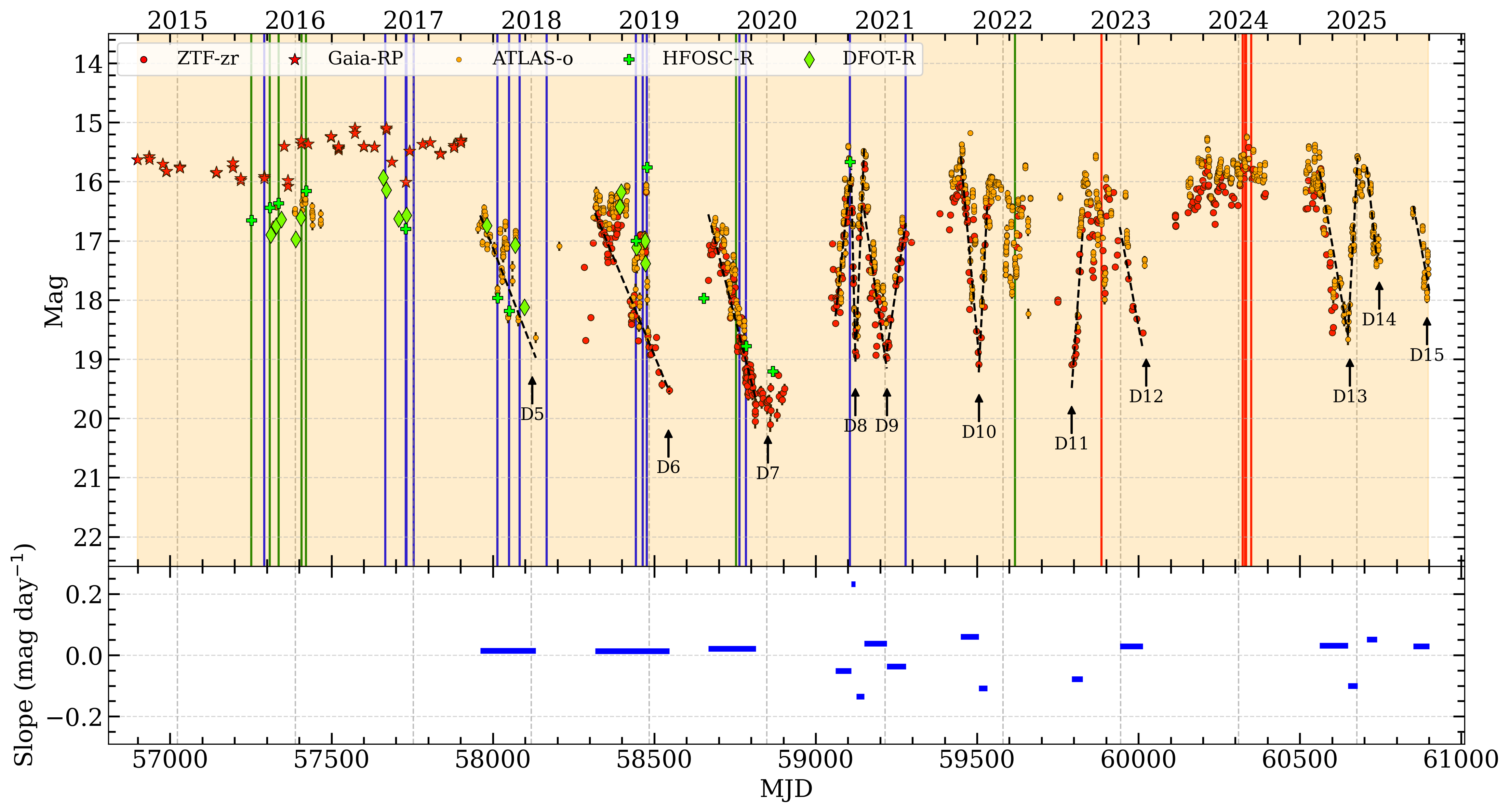}}
    \caption{\textit{Upper panel}: The historical $I$-band LC of V1180 Cas, covering the full range of previously published data from 1999 to 2022, is presented to highlight the source's long-term photometric variability. The data spanning 1999–2011 and 2011–2022 are taken from \citet{Kun_2011} and \citet{Mutafov_2022RAA....22l5014M}, respectively. The 7 dips reported in previous studies are marked using black arrows and labeled as D1, D2, and so on. The shaded region highlights our monitoring period.
    \textit{Lower panel}: The LC of V1180 Cas, combining catalog data and currently observed data. Catalog data includes ZTF , $zr$, Gaia DR3 $G_{RP}$, and ATLAS $o$. The current dataset consists of $R$-band observations obtained with HFOSC and DFOT. The vertical solid lines, colored blue, green, and red, represent the epochs of spectroscopic data, taken with HFOSC G7, HFOSC G8, and TANSPEC, respectively. The linear trends during fading and brightening events are marked using black dashed lines, and their slope (mag day$^{-1}$) values are shows in the lower sub-panel (for details, see Section \ref{sec:res_lc_ro}). All the large dips are marked with black arrows and labeled.
    In both panels, vertical dashed grey lines correspond to January 1st of each year, labeled on the upper x-axis. (The currently observed data used in this figure are available.)}
    \label{fig:lc_all}
\end{figure*}

The LCs of V1180 Cas, generated from all the aforementioned datasets, are presented in Figure \ref{fig:lc_all}.

\subsection{Spectroscopic data}\label{subsec:spectro}

We obtained optical spectroscopic data of V1180 Cas using HFOSC with Grism 7 (G7) and Grism 8 (G8) across 23 different epochs from 2015 to 2022 (see Table \ref{tab:obs_log}). G8 covers the spectral range 5800–9200 \r{A} with a resolving power of R$\sim$1200, while G7 spans 3800–8000 \r{A} with R$\sim$1000. Detailed information about the HFOSC instrument is available on the IAO website\footnote{\url{https://www.iiap.res.in/centers/iao/}}.

Additionally, we used TANSPEC to obtain optical–NIR spectra of V1180 Cas at eight different epochs since 2020 (see  Table \ref{tab:obs_log}), using its cross-dispersed (XD) mode. The XD mode of TANSPEC provides spectral coverage from 0.55 to 2.5 $\mu$m with a resolving power of R$\sim$1600 for a 1.0$^{\prime \prime}$ slit \citep{Sharma_2022PASP..134h5002S}.

The spectroscopic data obtained from HFOSC were preprocessed and reduced following the procedure described in \cite{Ghosh_2022ApJ...926...68G, Ghosh_2023ApJ...954...82G}. We employed the tasks \texttt{apall}, \texttt{identify}, and \texttt{dispcor} tasks available in Image Reduction and Analysis Facility (IRAF; \citealt{iraf1_1986SPIE..627..733T, iraf2_1993ASPC...52..173T}), to extract the spectra and perform wavelength calibration. For instrument response correction, we used the spectroscopic standard star Feige 34, observed on the same or adjacent nights. The instrument response function across the spectral range was derived by dividing the observed spectrum of the standard star by its reference spectrum from CALSPEC\footnote{\url{https://www.stsci.edu/hst/instrumentation/reference-data-for-calibration-and-tools/astronomical-catalogs/calspec}}.

For the TANSPEC spectroscopic data, spectral extraction was also carried out using the \texttt{apall} task in IRAF. In \texttt{multispec} mode, we utilized bright reference stars to trace the dispersion lines across 10 spectral orders in the 2D detector frame. These traced lines were then used to extract the spectra of V1180 Cas, accounting for any shifts in the cross-dispersion direction between the bright star and the target. Wavelength calibration and instrument response correction were performed using modules from the pyTANSPEC pipeline\footnote{\url{https://github.com/varghesereji/pyTANSPEC/blob/generalized_pipeline/doc/Pipeline_Documentation.md}} \citep{Ghosh_2023JApA...44...30G}.

Absolute flux calibration for the spectra from both instruments was carried out using contemporaneous or near-simultaneous photometric observations. For HFOSC optical spectra, we used photometric data in the $V$, $R$, and $I$ bands from \cite{Mutafov_2022RAA....22l5014M}, while for TANSPEC spectra, we used NIR $J$, $H$, and $K$ band photometry obtained with TANSPEC. The magnitudes were converted to fluxes using standard zero-point flux values for each filter. Scaling factors at the effective wavelengths of the filters were determined and fitted with a linear function, which was then extrapolated over the full spectral range to apply flux calibration.

\section{Results and Analysis} \label{sec:result}

In the following subsections, we present results from our long-term monitoring of V1180 Cas, incorporating both photometric (LCs) and spectroscopic observations.

\subsection{Photometric evolution of V1180 Cas} \label{subsec:lc}

Two major studies on V1180 Cas have been conducted by \citet{Kun_2011} and \citet{Mutafov_2022RAA....22l5014M} primarily focusing on its optical photometric evolution. In the following subsections, we briefly summarize their key findings and subsequently present results from more recent datasets. These include our own multi-epoch observations, along with an analysis of the long-term color evolution of V1180 Cas.

\subsubsection{Historical photometric evolution}

Figure \ref{fig:lc_all} shows the optical LC of V1180 Cas. It spans from 1999 to 2025. The photometric data from 1999 to 2011 are taken from \citet{Kun_2011}, and \citet{Mutafov_2022RAA....22l5014M} presented photometric data from 2011 to 2022.

\cite{Kun_2011} reported three large ($\Delta I \sim$3-5 mag) dips, which occurred in December 1999, March-April 2004, and January 2011, are marked as D1, D2, and D3 in the upper panel of Figure \ref{fig:lc_all}.
Later, \cite{Mutafov_2022RAA....22l5014M} identified four additional dips from 2011 to 2022: the D4 in September 2013, D5 in December 2017, D6 in February-March 2019, and D7 in January 2020. 
Further, two notable brightness increases were observed, the first peak in September 2020 and the second in July-August 2021 (see upper panel in Figure \ref{fig:lc_all}).

The largest ($\Delta I\sim$5 mag) dip (D2) lasted for about three years (2003–2006; \citealt{Kun_2011}). The second largest dip (D6) was $\Delta I \sim$4 mag, and the second longest dip (D3) lasted $\sim 1.5$ year (see the upper panel of Figure~\ref{fig:lc_all}). 

Based on optical and MIR LCs of V1180 Cas, \cite{Kun_2011} suggested that the observed variations are consistent with the model proposed by \citet{Turner_2010ApJ...708..188T}. In this model, dust clouds lifted into the disk atmosphere by magnetic activity cast moving shadows, causing brightness changes on timescales of days to months. However, changes in NIR colors and optical spectra indicate that changes in the accretion rate may also contribute to the star’s variability.

Based on $J$, $H$, $K$ photometry, \citet{Antoniucci_2014A&A...565L...7A} found that the NIR color evolution of V1180 Cas resembles that of EXors. Follow-up \textit{Chandra} X-ray observations suggested that a significant fraction of the photometric variability may be due to extinction rather than solely from accretion-driven excess emission \citep{Antoniucci_2015A&A...584A..21A}. Consistent with this interpretation, the optical LCs and color changes analysed by \citet{Mutafov_2022RAA....22l5014M} indicated that the variability in V1180 Cas is primarily driven by changes in extinction.

\subsubsection{Recent observations}
\label{sec:res_lc_ro}

The lower panel of Figure \ref{fig:lc_all} shows the LC of V1180 Cas from 2014 to 2024, compiled from survey data obtained from ZTF ($zg$, $zr$), Gaia DR3 ($G_{BP}$, $G$, $G_{RP}$) and ATLAS ($c$, $o$), together with our $R$ band observations from HFOSC and DFOT.

Compared to previously published LCs, the current dataset provides improved continuity and coverage.
Between August 2014 and May 2017, Gaia DR3 $G_{RP}$ data indicate that the star remained at a brightness level of $\sim$15.5 mag, with only modest fluctuations of $\lesssim$1 mag, consistent with typical T Tauri–like variability \citep{Grankin_2007A&A...461..183G, Audard_2014prpl.conf..387A}. This is also consistent with the observation of \citet{Mutafov_2022RAA....22l5014M}. From June 2017 to April 2023, the LC reveals relatively frequent, high-amplitude dips ($\Delta zr \sim$3.5 mag), occurring on timescales of about a year or less. 
Eleven prominent dips (D5-D15) observed during this period are marked with black arrows in the lower panel of Figure \ref{fig:lc_all}. In the initial three dips (D5, D6, and D7), only the fading phases are well-covered, with limited data capturing the brightening. In contrast, both the fading and brightening phases of the subsequent dips are well-sampled. Linear trends fitted to these fading and brightening phases are represented by black dashed lines, with corresponding slope values shown in the bottom panel of Figure~\ref{fig:lc_all}. A detailed discussion of these dips follows.

\begin{figure}
    \centering
    \subfigure{\includegraphics[width=1\linewidth]{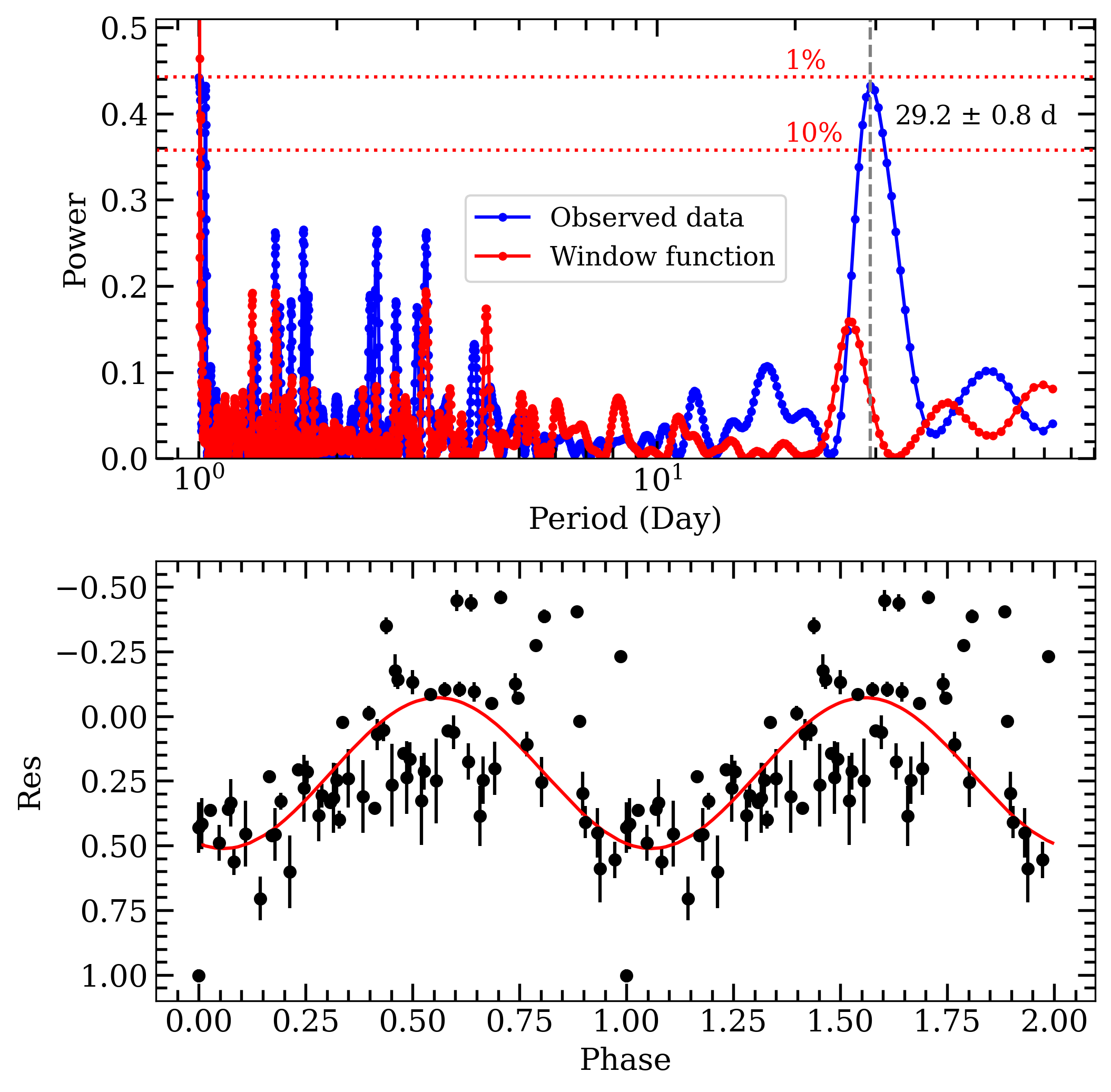}}
    \caption{The ZTF $zr$ band data, featuring high cadence (2-3 days) and regular observations between MJD 58665 and 58820 (D7), displays periodic variability in the LC.  
    The \textit{upper panel} illustrates the power spectrum for data (blue) and window function (red), obtained using a GLS periodogram. The red horizontal dotted lines indicate the 1$\%$ and 10$\%$ false alarm probability levels. The most significant peak is highlighted with a grey dashed vertical line. The \textit{lower panel} shows the phase-folded LC using a period of 29.2 days, which corresponds to the most prominent peak in the power spectrum. The red curve in the lower panel shows a sine curve with a period of 29.2 days.}
    \label{fig:period2_zr}
\end{figure}

\begin{itemize}
    \item {D5:} The first dip in the our LC appears around MJD 58138 (January 2018). The ATLAS $o$-band data show a linear fading from MJD 57958 to 58138 (180 days), during which the brightness decreases from $\sim$16.5 to 19.0 mag, corresponding to a fading rate of $\sim+0.014$ mag day$^{-1}$. Superimposed on this linear fading, we observe a sharp dip between MJD 57990–58005 (centered around MJD 57997) with an amplitude of $\sim$2 mag in $o$-band, followed by a $\sim$0.5 mag brightening event around MJD 58045. Additionally, stochastic variability with amplitudes ranging from a few millimagnitudes (mmag) to several hundred mmag is also present over timescales of a few days. The star returned to $\sim$16.5 mag around MJD 58318, indicating a duty cycle of roughly one year.
    \item {D6:} The second dip in the LC occurs around MJD 58547 (March 2019). From MJD 58318 to 58547, the LC shows a steady linear decline of $\sim$3 mag (from $\sim$16.5 to $\sim$19.5), with a slope of $\sim +0.013$ mag day$^{-1}$. This dip is very similar to D5. This decline is accompanied by brightening, with amplitudes reaching up to $zr$ $\sim$1.0 mag over timescales of a few days, particularly around MJD 58377 (September 2018) and MJD 58458 (December 2018). Due to a lack of data, the exact brightening rate could not be determined. However, it took approximately 138 days to reach $\sim$17 mag again, suggesting a slightly faster recovery rate of $\sim-0.018$ mag day$^{-1}$. Altogether, the full cycle---from $\sim$16.5 mag to $\sim$19.5 mag and back to $\sim$17 mag--- spans approximately one year.
    \item {D7:} The third dip in the LC occurs around MJD 58832 (December 2019). From MJD 58690 to 58814 (124 days), the LC shows a linear decline from $\sim$17.0 mag to $\sim$19.5 mag with a slope of approximately $+0.021$ mag day$^{-1}$, in the $zr$ band. Although data points are not available from MJD 58904, observations from \cite{Mutafov_2022RAA....22l5014M} confirm that V1180 Cas remained in a faint state for approximately 5–6 months, exhibiting only small-scale fluctuations. It regained its brightness level of 15.8 mag around MJD 59107 (September 2020) after $\sim$1.2 years.
The ZTF $zr$ band data, with a high cadence between MJD 58665 and 58820, exhibit periodic variability in the LC. To investigate, we applied a Generalized Lomb-Scargle (GLS; \citealt{Zechmeister_2018ascl.soft07019Z}) periodogram after subtracting the 0.021 mag day$^{-1}$ linear trend. Figure~\ref{fig:period2_zr} shows the resulting power spectrum (blue) along with the window function (red). False alarm probability levels of 1\% and 10\% were derived via bootstrap resampling (10,000 iterations). The strongest peak occurs at $29.2 \pm 0.8$ days, close to the synodic lunar period ($\sim$29.5 days). The Moon’s illumination was $\sim$20\% and $\sim$80\% during the phases corresponding to dip and peak in the folded LC, raising the possibility of lunar-related or cadence-induced systematics. Stellar rotation could, in principle, produce periodic variability from hot (accretion footprint) or cool (magnetic activity) spots, but typical periods are only a few days. Assuming Keplerian motion and neglecting the mass of the occulting structure---justified by the low disk-to-stellar mass ratios in Class II sources \citep{Beckwith_1990AJ.....99..924B, Beckwith_1996Natur.383..139B, Pascucci_2016ApJ...831..125P, Trapman_2025arXiv250610738T}---a 29.2-day period would correspond to occulting material orbiting at $\sim$0.16 au for a 0.65 $M_\odot$ \citep{Kun_2016ApJ...822...79K} star.
    
    \item {D8-D9:} Between September 2020 and February 2021, two consecutive dips are observed near MJD 59125 and 59218 (October 2020 and January 2021).
    From MJD 59057 to 59108, the brightness increases with a slope of $\sim-0.051$ mag day$^{-1}$, followed by a sharp decline at a rate of $\sim+0.232$ mag day$^{-1}$ until MJD 59125. Subsequently, the brightness increases sharply at a rate of $\sim-0.135$ mag day$^{-1}$ until MJD 59152. After this peak, the brightness decreases linearly by approximately 2.5 mag over 20 days. Beyond this, the LC exhibits a slower brightness decay at a rate of $\sim+0.024$ mag day$^{-1}$, with stochastic variations of a few hundred mmag, until MJD 59219. The brightness then increases linearly, recovering by approximately 2.3 mag over a span of two months. 
    Dip D8 starts from a magnitude level of $\sim$15.8 mag and returns to the same level in around 40 days. In D9, the star dimmed from $\sim$15.8 mag and returned to $\sim$16.8 mag in  $\sim$130 days.
    \item {D10:} The sixth major dip occurs around MJD 59505 (September 2021). Between MJD 59450 and 59505 (55 days), the $zr$-band brightness decreases linearly from 15.8 mag to 19.2 mag at a rate of $\sim+0.060$ mag day$^{-1}$, with a stochastic variation of amplitude $\sim$1.5 mag around MJD 59480. Subsequently, from MJD 59505 to MJD 59531 (26 days), the brightness increased at a rate of $\sim -0.108$ mag day$^{-1}$, reaching 16.5 mag. It took $\sim$80 days to complete this whole cycle. Between MJD 59531 and MJD 59740, the $o$ band data show additional stochastic variability up to 1.5 mag.
    \item {D11:} The seventh dip is observed around MJD 59793 \textbf{(}August 2022). Due to sparse data points between MJD 59540 and MJD 59793, the fading phase of the star is not captured in the LC. However, from MJD 59793 to MJD 59827 (34 days), the star's brightness increases by $\sim$2.5 mag, corresponding to a brightening rate of $\sim-0.078$ mag day$^{-1}$. Following this, the LC exhibits stochastic variability with an amplitude of $\sim$1 mag, continuing until MJD 59942. 
    \item {D12:} From MJD 59942 (December 2022) to MJD 60012 (March 2023), spanning a period of 70 days, the brightness decreases from $\sim$16.5 mag to $\sim$18.5 mag at a fading rate of $\sim+0.029$ mag day$^{-1}$. Sparse data points around MJD 60115, at a brightness level of $\sim$16.5 mag, suggest a recovery rate of $\sim-0.011$ mag day$^{-1}$ or possibly faster. From MJD 60115 to MJD 60212, the star's brightness slowly ($\sim -0.01$ mag day$^{-1}$) increases, followed by stochastic variability of amplitude $zr$ $\sim$1 mag till MJD 60392. A small dip ($\Delta zr\sim$1 mag) is also observed at around MJD 60237.
    \item {D13:} From MJD 60561 (September 2024) to 60649 (December 2024), the brightness decreases from $\sim$16 mag to $\sim$18.5 mag in the $o$ band, at a rate $\sim+0.031$ mag day$^{-1}$. It then recovered to 15.5 mag by MJD 60679, with a relatively higher rate of $\sim-0.1$ mag day$^{-1}$. The total duration of this dip is approximately 120 days.
    \item {D14-D15:} Between MJD 60708 (February 2025) to MJD 60898 (February 2025), data sampling is sparse. Nevertheless, two declines are identified: one from MJD 60708 to MJD 60739, and another from MJD 60852 to MJD 60898, with fading rates $\sim +0.05$ and $\sim -0.03$ respectively.
    
\end{itemize}



\begin{figure*}
   \centering
   \includegraphics[width=1\linewidth]{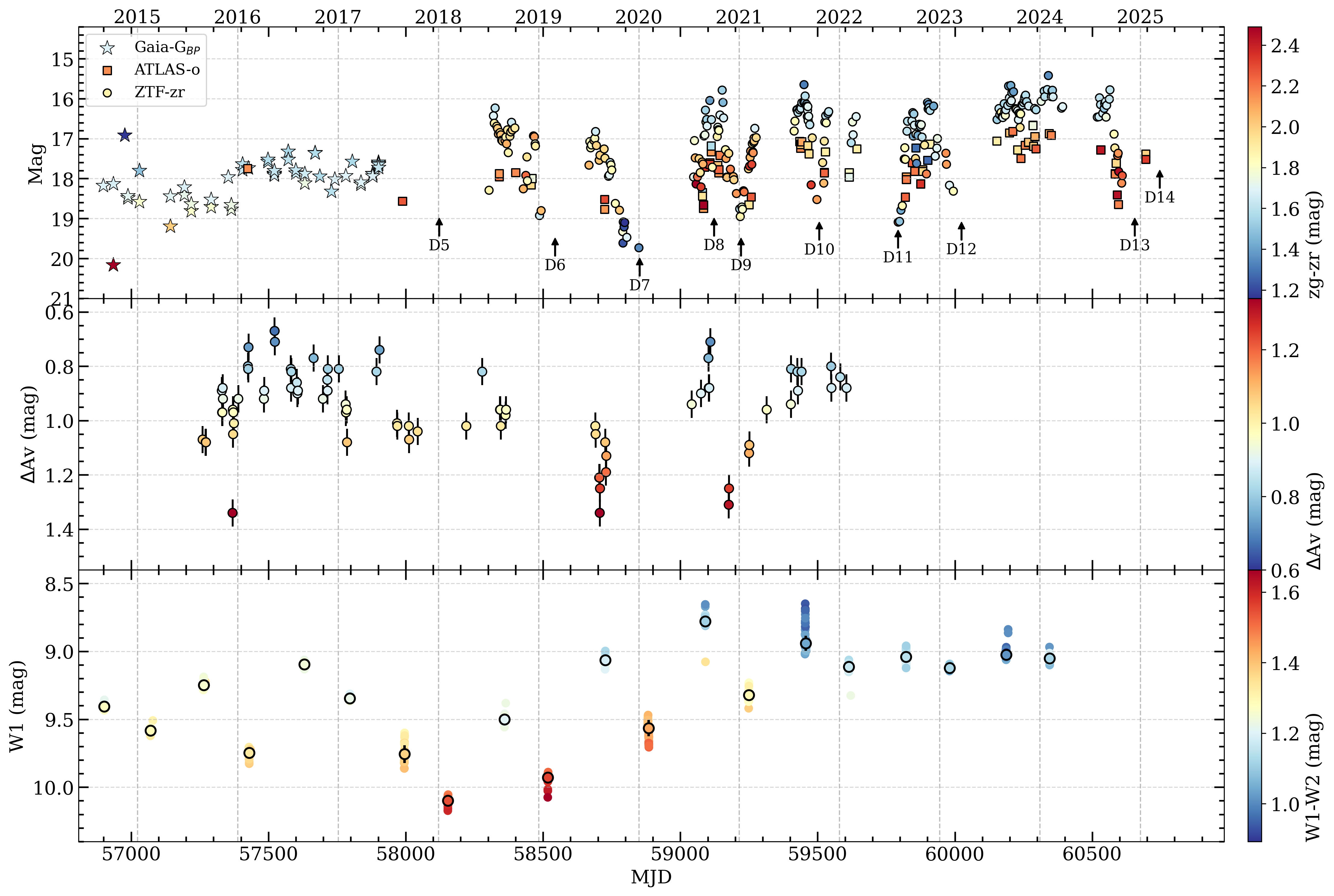}
   \caption{The optical, MIR color, and extinction ($A_V$) variation across different epochs. \textit{Upper panel}: Shows the optical color evolution of V1880 Cas using ZTF ($zg$, $zr$), ATLAS ($c$, $o$) and GAIA DR3 ($G_{BP}$, $G_{RP}$) data. Circles denote ZTF $zr$ mag color-coded by ($zg - zr$), star symbols represent Gaia $G_{RP}$ mag color-coded by ($G_{BP} - G_{RP}$), and squares indicate ATLAS $o$-band data color-coded by ($c - o$). The dips identified in the lower panel of Figure \ref{fig:lc_all} are also marked here. \textit{Middle panel}: The $A_V$ variation across different epochs with color coded using $\Delta A_V$ values. The $\Delta A_V$ values are calculated using color-color ($V-R$ versus $R-I$) diagram, using data from \cite{Mutafov_2022RAA....22l5014M}. \textit{Lower panel}: The LC using NEOWISE W1 data, color represents W1-W2 color. Black circles represent the median values of nearby observations (within 30 days).}
   \label{fig:lc_clr_evo}
\end{figure*}

The estimated brightening and dimming rates (in mag day$^{-1}$) during the large-amplitude dips in the LC of V1180 Cas using ATLAS $o$-band and ZTF $zr$-band data are shown in the lower sub-panel of Figure \ref{fig:lc_all}. These rates range from approximately $+0.013$ to $+0.232$ mag day$^{-1}$ for fading, and from $-0.037$ to $-0.135$ mag day$^{-1}$ for brightening. Notably, between MJD 59100 and MJD 59170, the rates reached hundreds of millimagnitudes (mmag) per day, whereas during other dips, they typically remained on the order of a few tens of mmag per day.

The temporal evolution of these rates reveals notable trends. For example, during the early prominent dip D2, both the fading and brightening rates were approximately $0.010$ mag day$^{-1}$. 
In the subsequent dips (D5–D6), the rise rate increased to $\sim-0.014$ and $-0.018$ mag day$^{-1}$, and in dips D7–D11, it varied between $-0.037$ (for D9) and $-0.135$ mag day$^{-1}$ (for D10).

\begin{figure*}
    \subfigure{\includegraphics[width=0.5\linewidth]{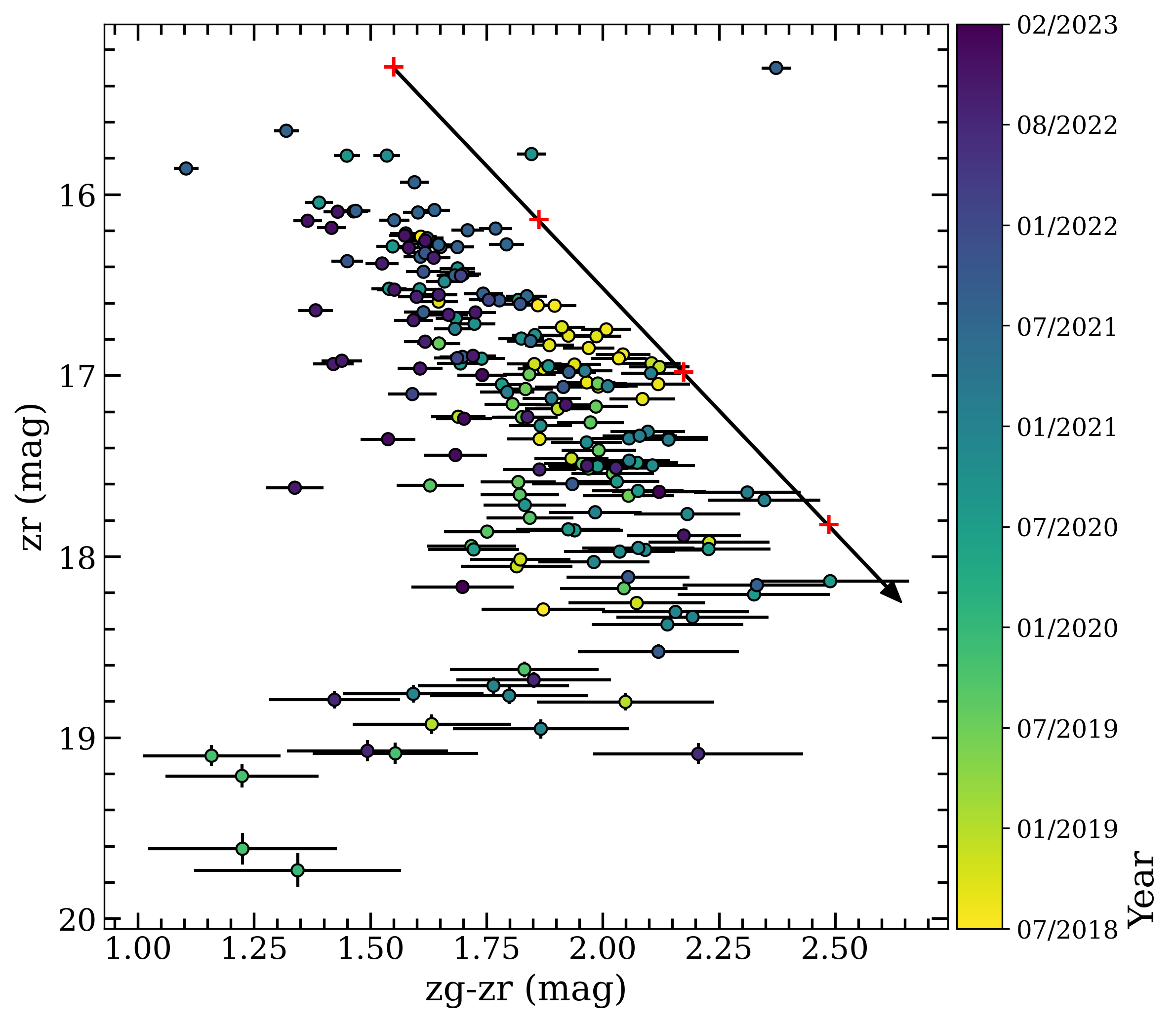}}
    \subfigure{\includegraphics[width=0.5\linewidth]{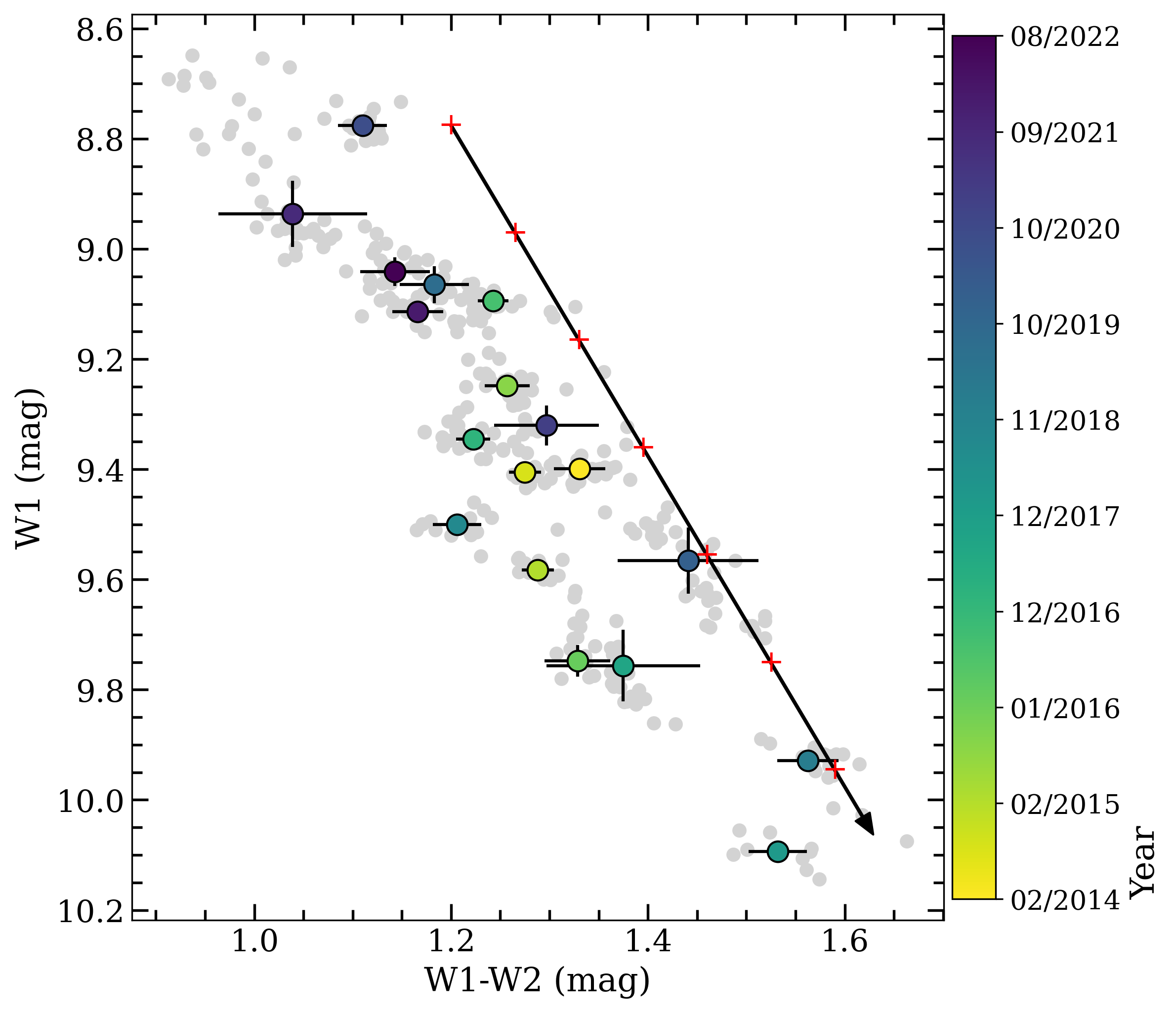}}
    \subfigure{\includegraphics[width=0.5\linewidth]{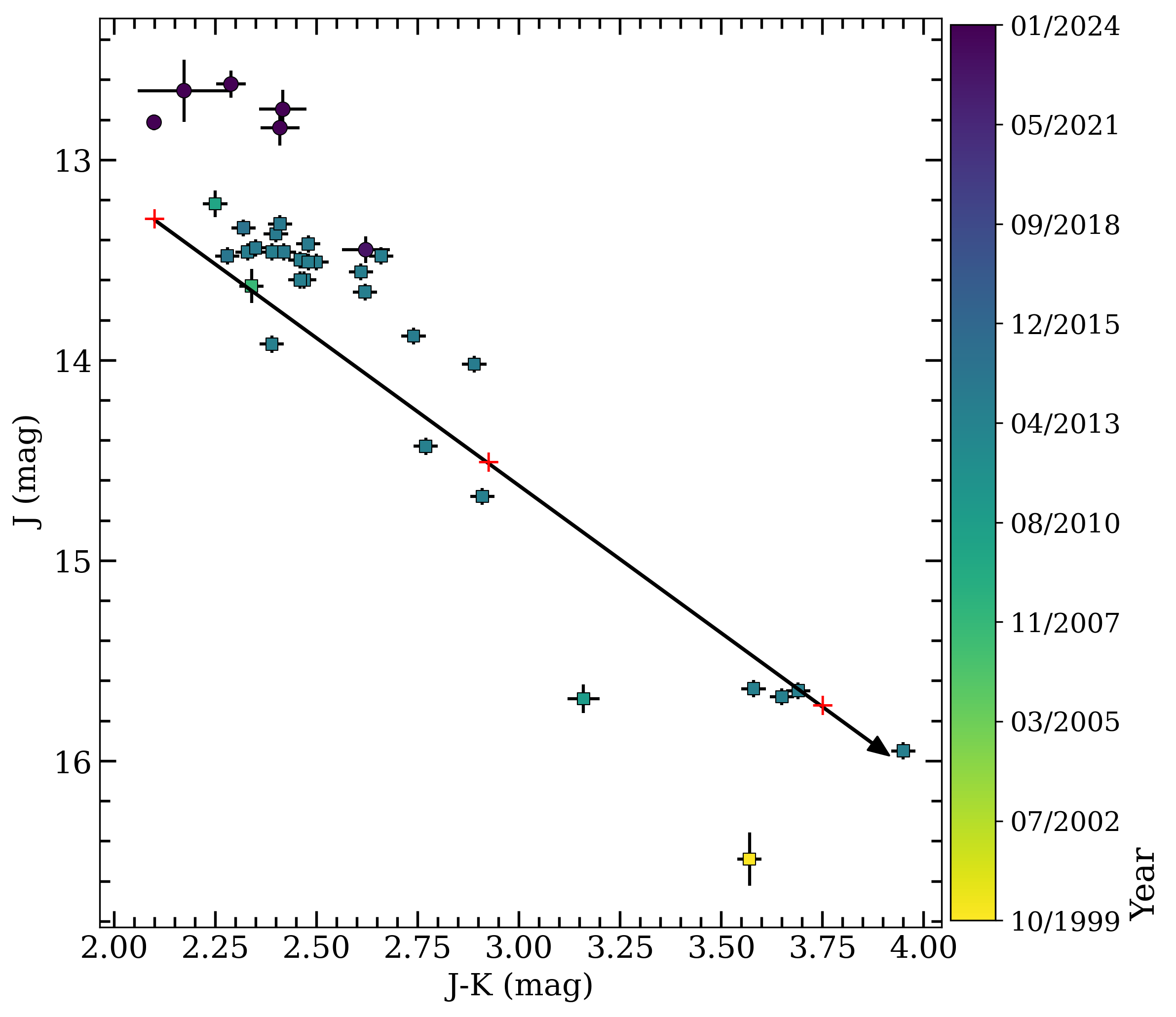}}
    \subfigure{\includegraphics[width=0.5\linewidth]{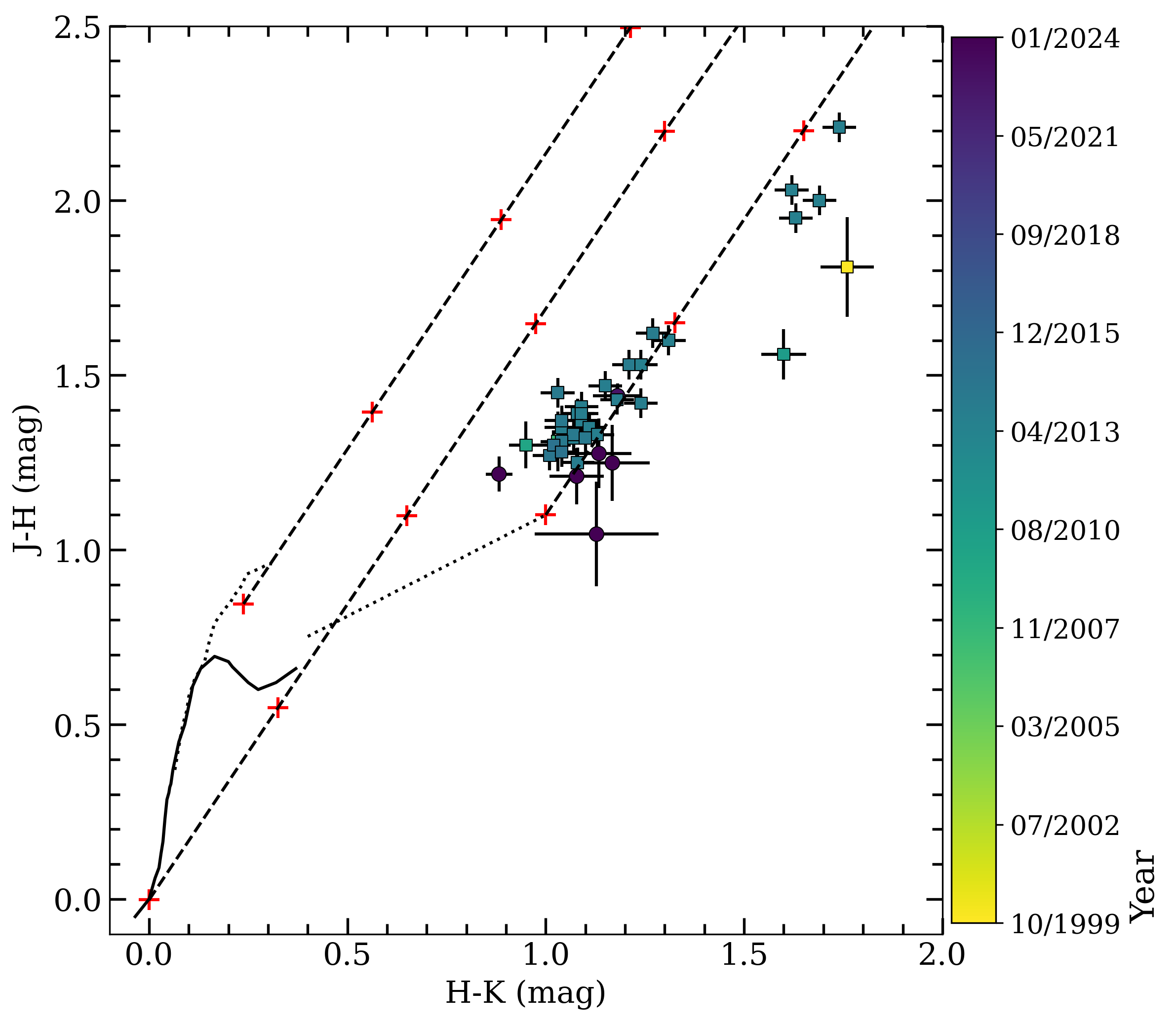}}
    \caption{ \textit{Upper panels}: CMD of V1180 Cas using ZTF ($zg$ and $zr$) and NEOWISE ($W1$ and $W2$) data, color represents observation epoch (in years) in both panels. The left panel shows $zr$ mag vs. ($zg-zr$) color. The right panel shows the MIR CMD, $W1$ vs. $W1$-$W2$. The grey data points are actual observations, while the black circles denote median values of nearby (within 30 days) observations. The black arrows in both panels show the extinction vector, and the red \enquote{+} sign denotes the $A_V$ difference of 1 mag in the optical CMD ($zr$ vs. $zr-zg$) and 5 mag in the MIR CMD ($W1$ vs. $W1$-$W2$). \textit{Lower panels}: The CMD (left) and color-color diagram (right) using $J$, $H$, and $K$ bands photometric data. The square data points represent the data from the literature \citep{Kun_2011, Antoniucci_2014A&A...565L...7A}, and the circles represent data observed from TANSPEC. In the left panel, the black arrow represents the extinction vector with marked 5 mag differences in $A_V$ as a red \enquote{+} sign. In the right panel, the solid curve shows the locus of field dwarfs, and the dotted curve shows the locus of giants \citep{Bessell_1988PASP..100.1134B}. The dotted line represents the locus of classical T-Tauri (CTT) stars \citep{Meyer_1997AJ....114..288M}. The diagonal straight dashed lines show the reddening vectors \citep{Rieke_1985ApJ...288..618R}, with red \enquote{+} signs denoting an $A_V$ difference of 5 mag. The color-code represents the observation epochs (MM/YYYY) labeled on the right side of the figures. (The TANSPEC data used in this figure are available.)
    }
    \label{fig:cmd_ztf}
\end{figure*}

\subsubsection{Color variation} \label{subsec:cc}

In Figure \ref{fig:lc_clr_evo} and Figure \ref{fig:cmd_ztf},  we compare the photometric variations in the LC with the color evolution of the star, both in optical and MIR bands.

\paragraph{Optical color}

The upper panel of Figure~\ref{fig:lc_clr_evo} shows the optical color evolution of V1880 Cas using ZTF ($zg$, $zr$), ATLAS ($c$, $o$) and GAIA DR3 ($G_{BP}$, $G_{RP}$) data.
The optical color evolution from 2018 to 2023 shows that V1180 Cas generally becomes redder as it fades. However, there are notable exceptions—such as in November–December 2019, January 2021, and August 2023—when it becomes bluer while fading. This behavior is further illustrated in the upper-left panel of Figure~\ref{fig:cmd_ztf}, which presents the optical color–magnitude diagram (CMD) based on ZTF data. For brighter data points (mag $<$ 18), the distribution aligns with the reddening vector (indicated by a black arrow). Toward fainter magnitudes, however, a clear color reversal—i.e., blueing when fading—is observed.

This \enquote{blueing effect} was also previously reported by \citet{Mutafov_2022RAA....22l5014M} using decade-long ($V$, $R$, and $I$ band) optical data spanning 2011–2022. Similar phenomena have been documented in other UXor-type stars such as V1184 Tau \citep{Semkov_2015PASA...32...11S}, GM Cep \citep{Mutafov_2021arXiv210506221M}, and V852 Aur \citep{Abraham_2018ApJ...853...28A}. The \enquote{blueing effect} in UXors arises due to increased contributions from scattered light: when orbiting circumstellar dust structures obscure the direct starlight but not the surrounding scattering regions, the observed flux becomes dominated by scattered (and intrinsically bluer) light \citep{Natta_2000A&A...364..633N}. Aside from the blueing episodes, the color evolution of V1180 Cas largely follows the extinction vector in the optical CMD, albeit with significant scatter. The resemblance of its color behavior to that of UXor-type stars strongly suggests a similar system configuration, involving variable circumstellar extinction and scattered light contributions.

The middle panel of Figure~\ref{fig:lc_clr_evo} presents the $A_V$ distribution, derived from the optical color–color ($V-R$ vs. $R-I$) diagram using photometric data from \citet{Mutafov_2022RAA....22l5014M}.  Prior to constructing the color–color (CC) diagram, we used the CMDs ($V$ vs. $V-I$ and $R$ vs. $R-I$; see Figure \ref{fig:av_clc} in Appendix) to identify and exclude data points exhibiting the \enquote{blueing effect}. The final CC diagram ($V-R$ vs. $R-I$) is shown in right panel of Figure \ref{fig:av_clc} in Appendix. To estimate extinction difference, we constructed reddening vectors for each data point in the ($V-R$ vs. $R-I$) CC diagram, originating from a reference line at $R-I = 0.65$ mag. These vectors follow the standard reddening law with $R_V = 3.1$, and the $\Delta A_V$ value for each point is computed as the vector's length in the CC diagram space.

A comparison between the extinction variations and the LC (top panel of Figure \ref{fig:lc_clr_evo}) reveals several broad trends. During the small-amplitude brightening ($\Delta I \sim 0.5$ mag) around September 2015 (MJD $\sim$57270), the $A_V$ values are slightly elevated ($\Delta A_V \sim 0.2–0.3$ mag) relative to those in 2016. However, these variations are comparable to the scatter in the data and should be regarded as marginal indications rather than statistically significant changes. Similarly, modest increases in $A_V$ at the onset of dips D5 and D6 are observed.
In contrast, more pronounced extinction variability is observed during dip D7. During the declining phase, $A_V$ reaches its highest values ($\Delta A_V \sim 1.3$ mag) at intermediate brightness levels, while near the center of the dip, the extinction decreases to $\Delta A_V \sim 1.1$ mag. As the source recovers from D7, $A_V$ returns to values comparable to those measured in 2016. A similar, statistically significant increase in extinction is observed during the fading phase of D9, followed by a gradual decrease in $A_V$ as the source brightens.
For several other dips (D8, D10–D15), the limited availability of CC measurements precludes a detailed analysis of the extinction behavior.

Overall, the extinction $A_V$ tends to increase during the onset or intermediate phases of photometric dips, indicating enhanced circumstellar obscuration. In contrast, at the deepest stages of the dips (notably in D7), a reduction in $A_V$ is observed, consistent with the \enquote{blueing effect} and suggesting an increased contribution from scattered light when the direct stellar emission is heavily obscured.

\begin{figure*}
    \centering

    \includegraphics[width=1\linewidth]{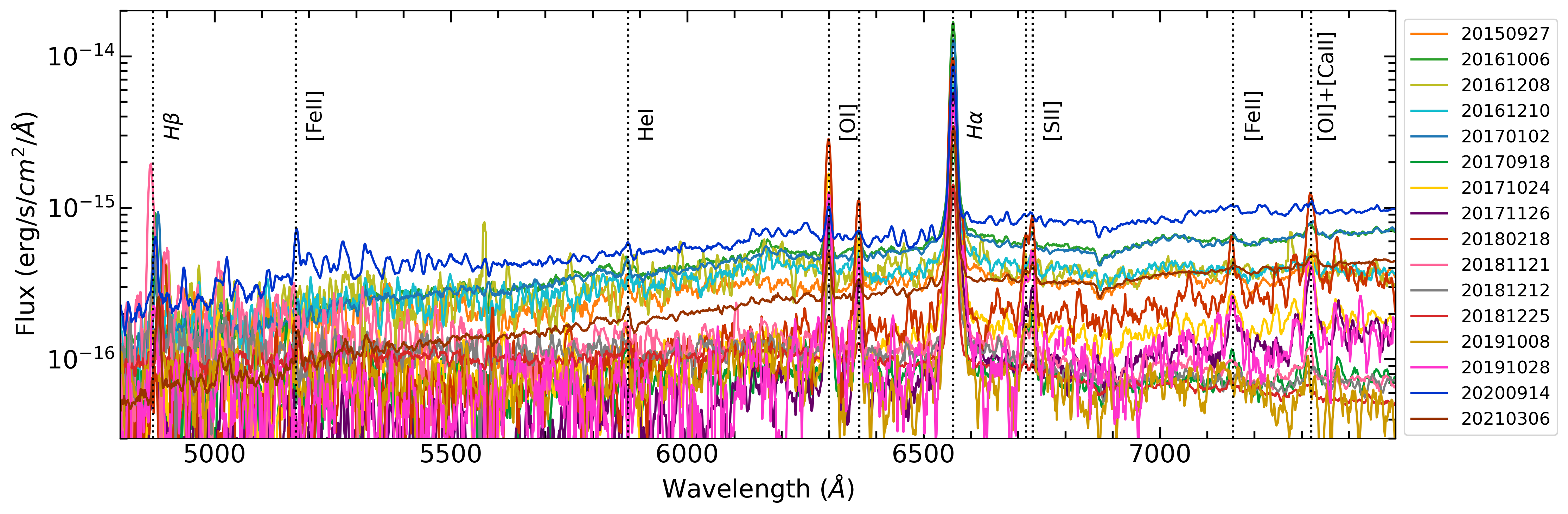}
    \includegraphics[width=1\linewidth]{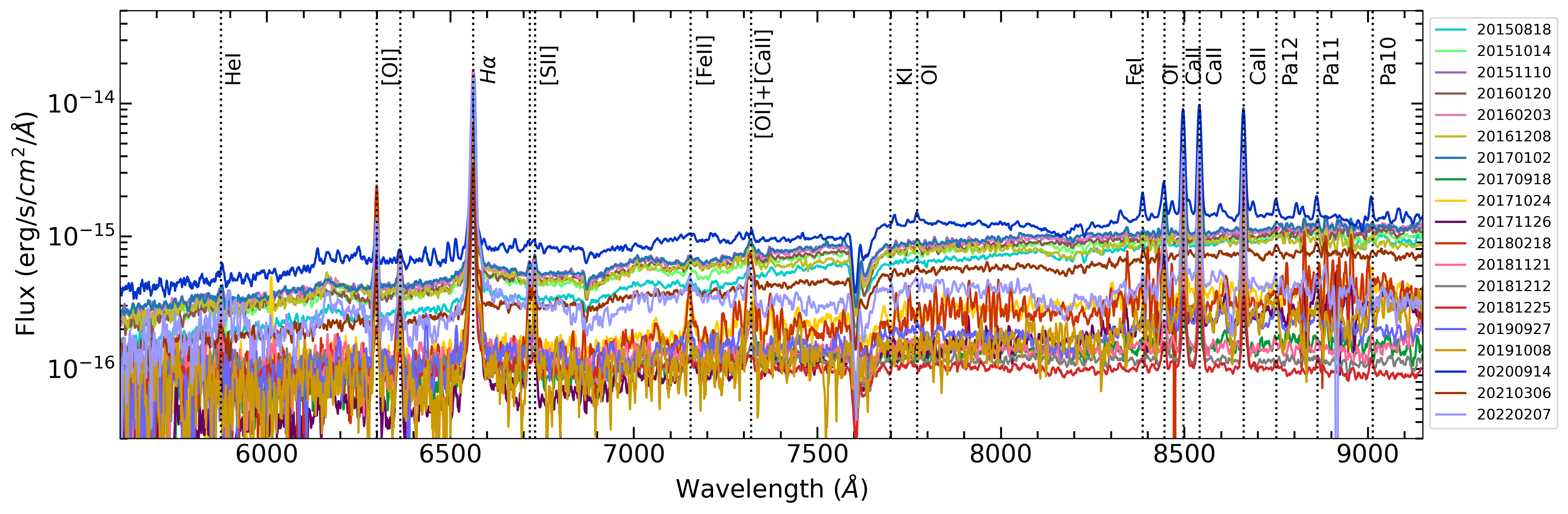}
    \caption{Flux-calibrated optical spectra of V1180 Cas. The \textit{upper} and \textit{lower} panels display spectra obtained with the HFOSC instrument using the G7 and G8 grisms, respectively. Spectra from different epochs are shown in various colors, labeled in the format YYYYMMDD. Several emission lines are indicated by dotted black lines and labeled.}
    \label{fig:spec_hfosc}
\end{figure*}

\begin{figure*}
    \centering
    \includegraphics[width=1\linewidth]{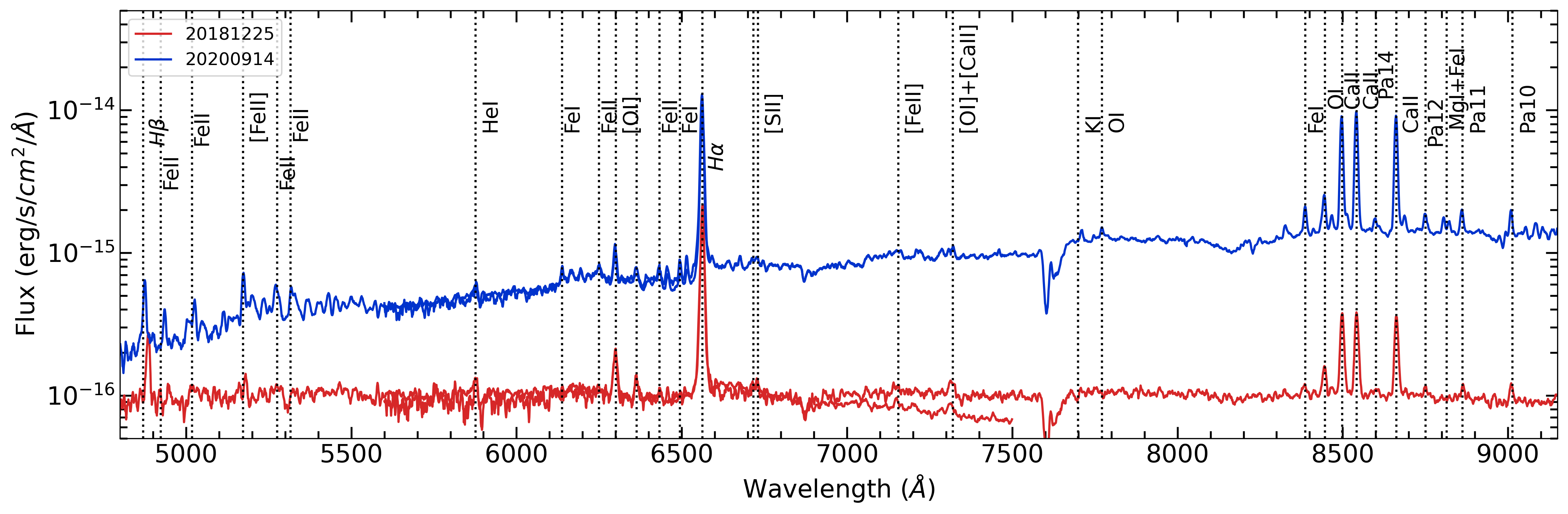}
    \caption{Comparison of the optical spectrum taken in the brightest phase (20200914) and in the faintest phase (20181225). In both epochs, the optical spectra were taken using both grisms G7 and G8 in HFOSC; this figure includes both spectra taken using G7 and G8. The prominent lines are marked with a black dashed line and labeled with their names.}
    \label{fig:spec_hfosc_comp}
\end{figure*}

\begin{figure*}
    \centering
    \includegraphics[width=1\linewidth]{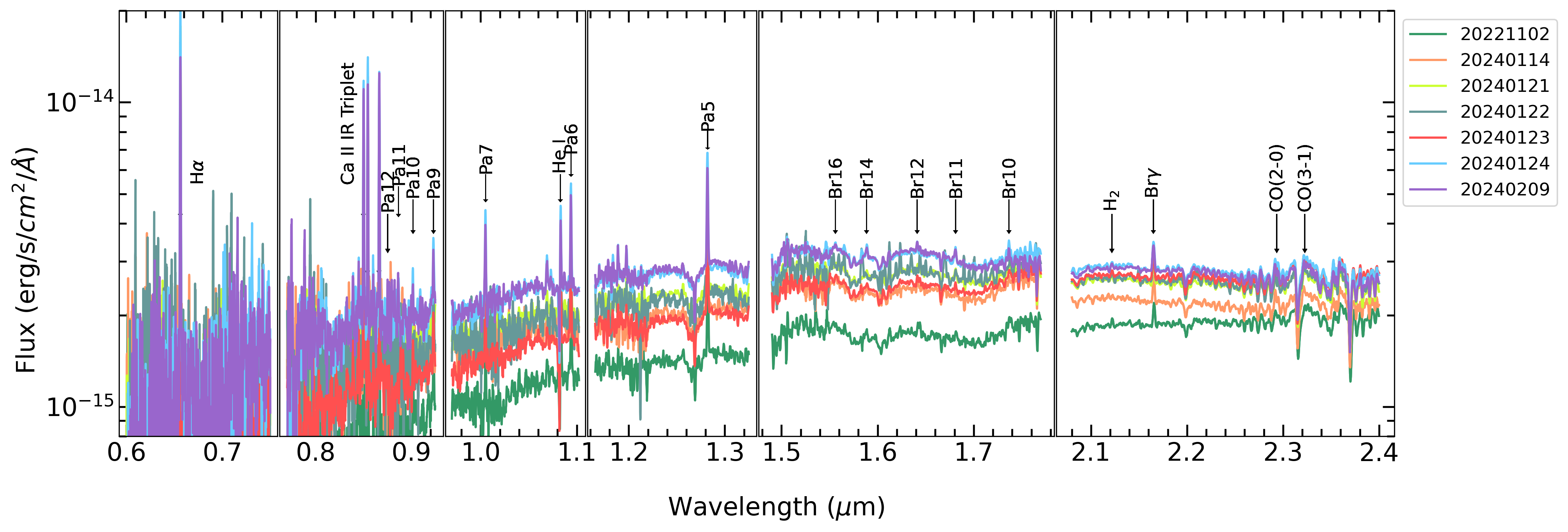}  
     \includegraphics[width=0.97\linewidth]{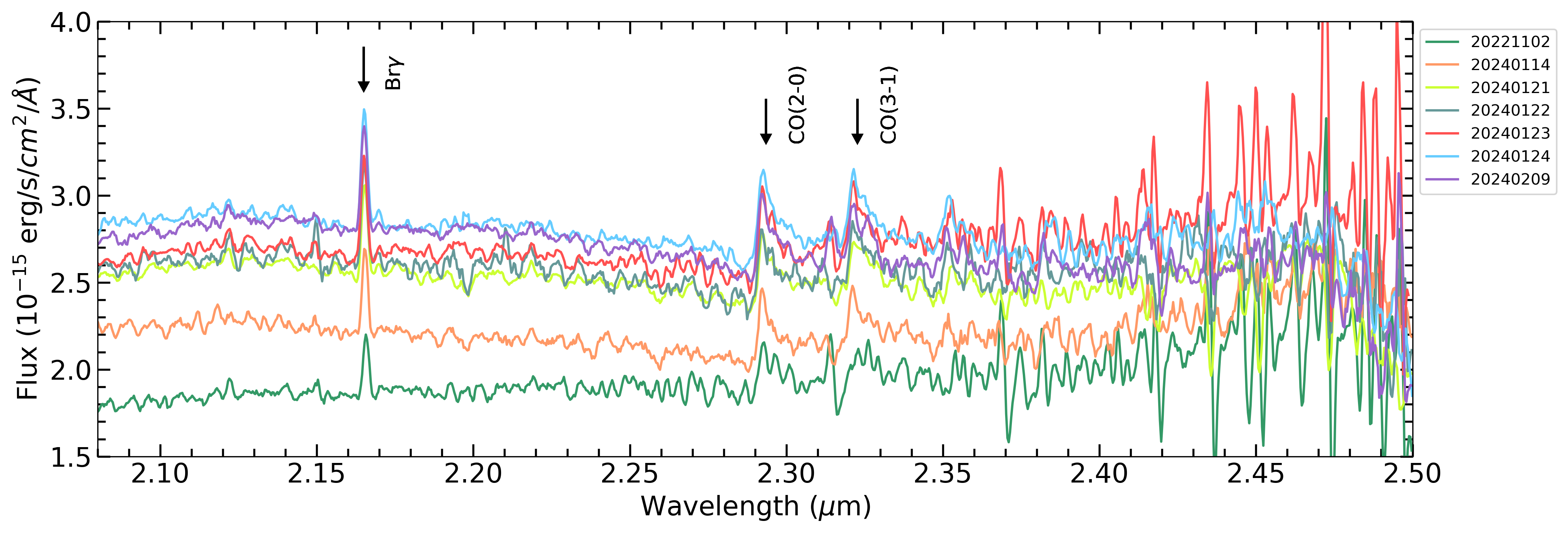}
    \caption{\textit{Upper panel}: Flux-calibrated optical-NIR spectra of V1180 Cas taken using TANSPEC. The identified lines are labeled and marked with black arrows. The regions affected by atmospheric absorption windows are excluded from the spectrum. 
    \textit{Lower panel}: Telluric corrected K band TANSPEC spectra of V1180 Cas. The locations of bandheads CO (2-0) and CO (3-1) are marked, and we see significant emission at the location of these lines.
    }
    \label{fig:tanspec_spec}
\end{figure*}

\paragraph{NIR/MIR colors}

The lower panel of Figure~\ref{fig:lc_clr_evo} presents the MIR LC of V1180 Cas, constructed using NEOWISE $W1$-band photometric data, with data points color-coded according to their corresponding $W1 - W2$ values. The overall morphology of the MIR LC closely mirrors that of the optical. Although we could not perform any measurements of time delay between the optical and MIR LC due to the sparse cadence of the WISE data, WISE band shows a concurrent decrease in brightness during the dips in the optical LC.

Between MJD 57930 and MJD 59460—spanning dips D5 through D9—the MIR LC exhibits a clear quasi-periodicity of approximately one year. However, the amplitude of variability in the MIR is notably smaller ($\sim$0.5-1 mag in the $W1$ band) than in the optical LC ($\sim$2.5-3.0 mag). Starting around MJD 57930, where the first prominent dip D5 is observed, the MIR LC shows two data points per cycle: one representing the bright phase and the other the faint phase. This alternating pattern results in a zig-zag structure, reflecting a similar periodic modulation seen in the optical LC.

Superimposed on this periodicity is a gradual, long-term brightening trend in the MIR LC. This secular increase in brightness may suggest an evolving inner disk emission, possibly due to changes in disk temperature or accretion-related heating.

During the fading events—particularly dips D5, D6, D7, and D9—the MIR LC exhibits significant reddening, as seen from the color-coded $W1-W2$ values. This behavior contrasts with the optical LC, where blueing is sometimes observed during fading events. This discrepancy is further illustrated in the upper-right panel of Figure~\ref{fig:cmd_ztf}, which shows the MIR CMD based on WISE data. The MIR CMD supports the interpretation that the star becomes redder when fainter, consistent with variable extinction due to circumstellar material.

The lower panels of Figure~\ref{fig:cmd_ztf} present the NIR CMD and CC diagram, constructed using $J$, $H$, and $K$-band data from both the literature (square symbols) and TANSPEC observations (solid circles). 
The NIR CMD clearly shows that V1180 Cas becomes redder as it dims, further supporting the extinction-driven variability interpretation. 
Further, as can be seen in the NIR CC diagram, during the bright states ($J <$ 15.5 mag), the star's position aligns well with the classical T Tauri locus. In contrast, during faint states ($J \geq$ 15.5 mag), the source shifts toward redder colors, but with a noticeable IR excess. Similar behavior has been observed in eruptive YSOs such as HBC 722 and VSX J205126.1+440523 \citep{Kospal_2011A&A...527A.133K}.
This departure from the reddening vector suggests that, in addition to extinction, the observed variability may also involve cooling of the continuum emission or structural changes in the inner disk, possibly due to changes in accretion or reprocessing of stellar radiation \citep{Meyer_1997AJ....114..288M}.

\subsection{Spectra}\label{subsec:res_spec}

As V1180 Cas exhibits large-amplitude variability across multiple timescales, it represents a highly dynamic system that warrants continued spectroscopic follow-up to constrain the physical processes driving its behavior. Given our extensive spectroscopic monitoring of the source over the past decade—spanning both optical and NIR wavelengths—we now present a detailed discussion of these results in the following section.

\subsubsection{HFOSC optical spectra}

The optical spectra were acquired over 23 epochs between 2015 and 2021 using the HFOSC instrument on the HCT. Figure \ref{fig:spec_hfosc} presents the flux-calibrated optical spectra of V1180 Cas observed with the grism G7 (lower panel) and G8 (upper panel).
The different colors represent the different epochs on which the spectra were taken, labeled (using the format YYYYMMDD) on the right side of the figures. The prominent lines in the spectra are marked using black dotted lines and labeled with their names.

The optical spectra of V1180 Cas shows strong H${\alpha}$ ($6563$ \r{A}) emission, Ca II IR triplet (8498 \r{A}, 8542 \r{A} and 8662 \r{A}), forbidden [O I] (6300 \r{A}), and H${\beta}$ (4861 \r{A}) lines in emission. Additionally, several weaker lines are visible, including [O I] (6363 \r{A}), [S II] (6717, 6731 \r{A}), He I (5876 \r{A}), blended [O I] and [Ca II] at 7320 \r{A}, O I (7772, 8446 \r{A}), many Fe I, Fe II and [Fe II] lines and Paschen (Pa) 12, 11, 10 at (8750, 8863, 9013 \r{A}) (see, Figure \ref{fig:spec_hfosc}).
The presence of hydrogen recombination lines and the Ca II IR triplet in emission indicates ongoing magnetospheric accretion in the YSO V1180 Cas, while the presence of forbidden lines such as [O I] and [S II] suggests an association with outflow activity \citep{Muzerolle_1998AJ....116..455M, White_2003ApJ...582.1109W, Alcala_2014A&A...561A...2A, Ninan_2015ApJ...815....4N}.  
 
The flux-calibrated optical spectra (see Figure \ref{fig:spec_hfosc}) show variations in the spectral continuum slope that correspond to changes in photometric brightness. This behaviour is expected, as the spectra are flux-calibrated using contemporaneous or near-simultaneous photometric data. During the bright state (see Figure \ref{fig:spec_hfosc_comp}), the continuum displays a steeper slope with enhanced emission in the red part of the spectrum and an overall higher flux level. In contrast, during the dip in the LC, the continuum exhibits a relatively flattened slope and a lower flux level. Several emission features (e.g. Ca II IR triplet, Pa10, Pa11, Pa12 and Fe I) exhibit larger line fluxes during the bright state as compared to the dip in the LC. Additionally, the K I and O I lines at 7698.96 \r{A} and 7771.94 \r{A}, respectively, are seen in emission during the bright state, while during the faint state these lines are absent in the optical spectra of V1180 Cas (see Figure \ref{fig:spec_hfosc_comp}). 

\begin{figure*}
    \centering
    \includegraphics[width=1\textwidth]{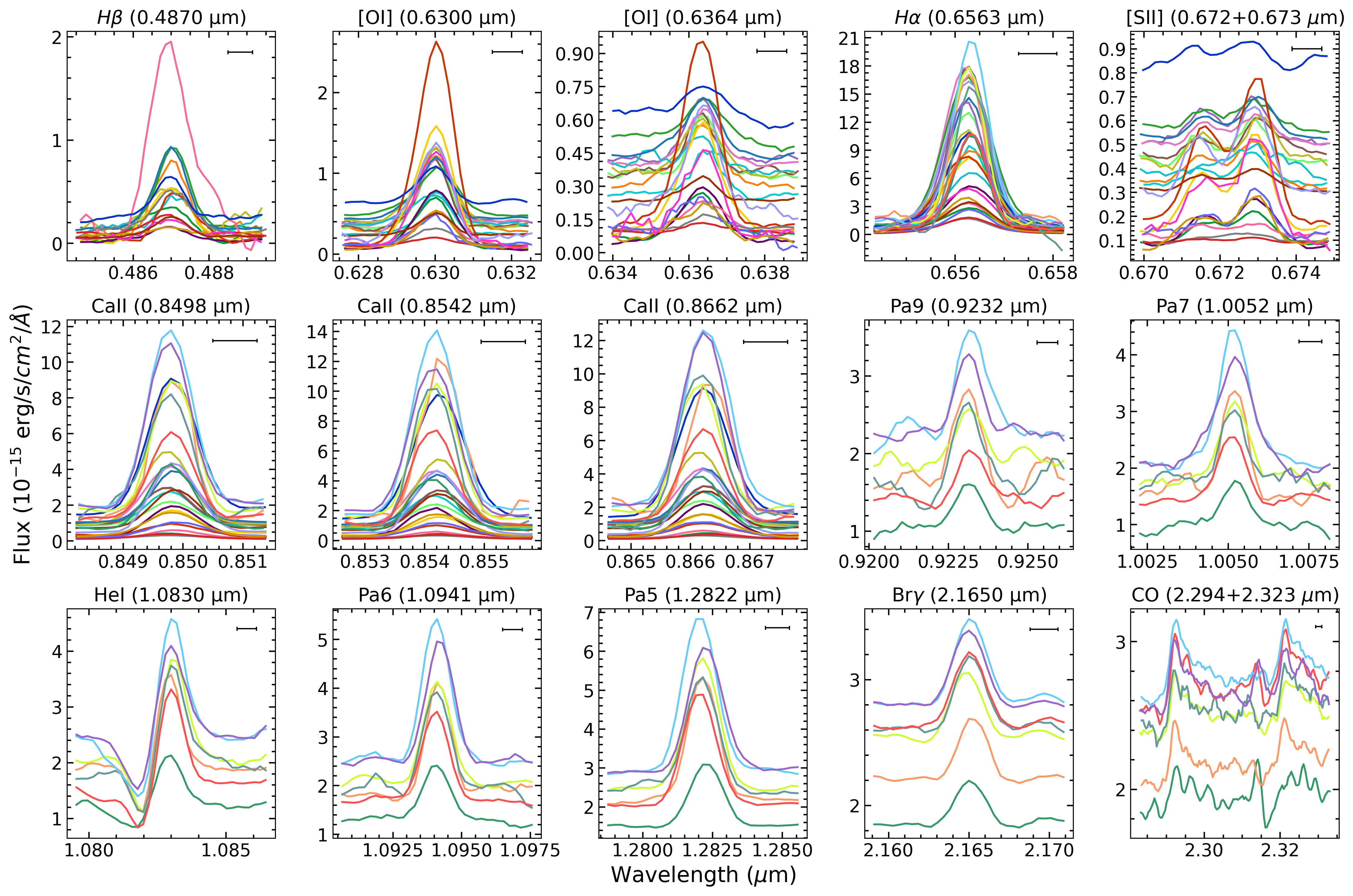}
    \caption{The temporal evolution of prominent spectral lines identified in HFOSC/TANSPEC spectra of V1180 Cas. The name of each line is given as the title of the subplots. The colors represent the different epochs as labeled in Figures \ref{fig:spec_hfosc}, and \ref{fig:tanspec_spec}. The black horizontal bars in the upper-right corner of each subplot indicate the instrumental line broadening at the central wavelength of the line.}
    \label{fig:tanspec_cutouts}
\end{figure*}

\subsubsection{TANSPEC optical-NIR spectra}

Figure \ref{fig:tanspec_spec} presents the optical to NIR spectra of V1180 Cas, covering the wavelength range $\sim$0.6–2.5~$\mu$m, obtained with TANSPEC over seven epochs between 2022 and 2024.
 
Although the optical portion suffers from a low signal-to-noise ratio, several key features remain identifiable, including H${\alpha}$, the Ca II IR triplet, and Paschen series emission lines. The noticeable variations in the continuum emission are observed over a timescale of several weeks.
The NIR spectrum of V1180 Cas is dominated by emission features, except for He I at 1.083 $\mu$m, which exhibits a P-Cygni profile. Detected lines include hydrogen recombination transitions from the Paschen series (Pa5–Pa12), Brackett series (Br10–Br16 and Br${\gamma}$). A weak H$_{2}$ emission line is also detected at 2.1218 $\mu$m.
We also observe significant emission lines of CO (2–0), and CO (3–1) band-heads at 2.2935 and 2.3227 $\mu$m, respectively.

\subsubsection{Spectral line profiles and their evolution}

Figure~\ref{fig:tanspec_cutouts} presents the line profiles of several prominent spectral transitions observed at different epochs using both HFOSC and TANSPEC data. These profiles provide an opportunity to investigate temporal variations in line strength, morphology, and kinematics, offering crucial insights into the accretion and outflow processes in V1180 Cas.

Emission features detected in these spectra serve as vital diagnostics for probing both accretion and outflow mechanisms. For instance, hydrogen recombination lines primarily trace accretion activity, with their luminosities and profiles indicative of the mass accretion rate and the structure of the magnetospheric accretion columns. Forbidden lines like [O I], [SII], and [Fe II], on the other hand, are classical signatures of outflows and shocks, revealing the presence and variability of jets and disk winds \citep{Hamann_1994ApJS...93..485H, Whelan_2004A&A...417..247W, Fang_2018ApJ...868...28F, Whelan_2024ApJ...974..293W}. 

Each line also originates from a distinct region of the circumstellar environment. The H$\alpha$ line, for example, is known to form in hot accretion spots on the stellar surface and within the magnetospheric accretion columns. It can also include contributions from chromospheric activity and disk winds \citep{Natta_2004A&A...424..603N, Fang_2013ApJS..207....5F, Alcala_2014A&A...561A...2A, Ninan_2015ApJ...815....4N}. The Ca II IR triplet arises either from optically thick magnetospheric regions—where the equivalent width (EW) ratios are close to 1:1:1—or from optically thin outflows, which exhibit an EW ratio of roughly 1:9:5 \citep{Hamann_1992ApJS...82..247H, Fernandez_2001A&A...380..264F, Alcala_2014A&A...561A...2A}. Paschen and Brackett series lines, including Br$\gamma$, are commonly formed in the magnetospheric infall zones \citep{Muzerolle_2001ApJ...550..944M, Alcala_2014A&A...561A...2A}.
The He I 1.083 $\mu$m line is a particularly important tracer, as it may carry signatures of both accretion and outflow activity. Its blue-shifted absorption component, when present, is a direct indicator of outflowing winds \citep{Edwards_2006ApJ...646..319E, Alcala_2014A&A...561A...2A}. Meanwhile, CO overtone bandhead emission, arising from the outer parts of the inner disk, indicates thermal excitation by irradiation from the central star---often enhanced during episodes of elevated accretion \citep{Calvet_1991ApJ...383..752C}.

The current spectral sample shows that the line intensities vary substantially across different epochs. However, the overall line profiles remain relatively stable. This suggests that the geometry and origin of the emitting regions remain largely unchanged, even though the activity level fluctuates. The simultaneous presence of hydrogen recombination lines (H$\alpha$, Br$\gamma$, Paschen and Brackett series) and CO bandhead emission points to ongoing magnetospheric accretion from the inner disk onto the star. The observed variability in their line strengths likely reflects changes in the mass accretion rate. At the same time, the detection of blue-shifted absorption in the He I line, along with forbidden transitions such as [O I] (6300, 6363 \r{A}), [S II] (6717, 6731 \r{A}), the blended [O I]+[Ca II] complex near 7320 \r{A}, and [Fe II], confirms the presence of strong outflows. The persistence of these outflow tracers across all observed epochs suggests that the outflow is a continuous process in V1180 Cas, albeit with variable strength, likely modulated by episodic accretion events.

It is noteworthy that while the He I line clearly exhibits a P-Cygni profile, such a feature is absent in the H$\alpha$ line. This contrast suggests that the He I line is more effective in tracing inner, line-of-sight wind components, likely due to its sensitivity to lower-density, high-velocity outflows and its formation in a metastable level conducive to absorption. In contrast, the H$\alpha$ line is often dominated by strong accretion-related emission and may form in regions or geometries less favorable for producing detectable absorption features. In addition, the limited spectral resolution of the observations may further hinder the detection of subtle profile changes or a weak P-Cygni component in H$\alpha$. Together, these factors highlight  the complex interplay between accretion and wind processes in V1180 Cas and underscore the importance of using multiple diagnostics to probe different components of the circumstellar environment.

\begin{figure*}
    \centering
     \subfigure{\includegraphics[width=0.49\linewidth]{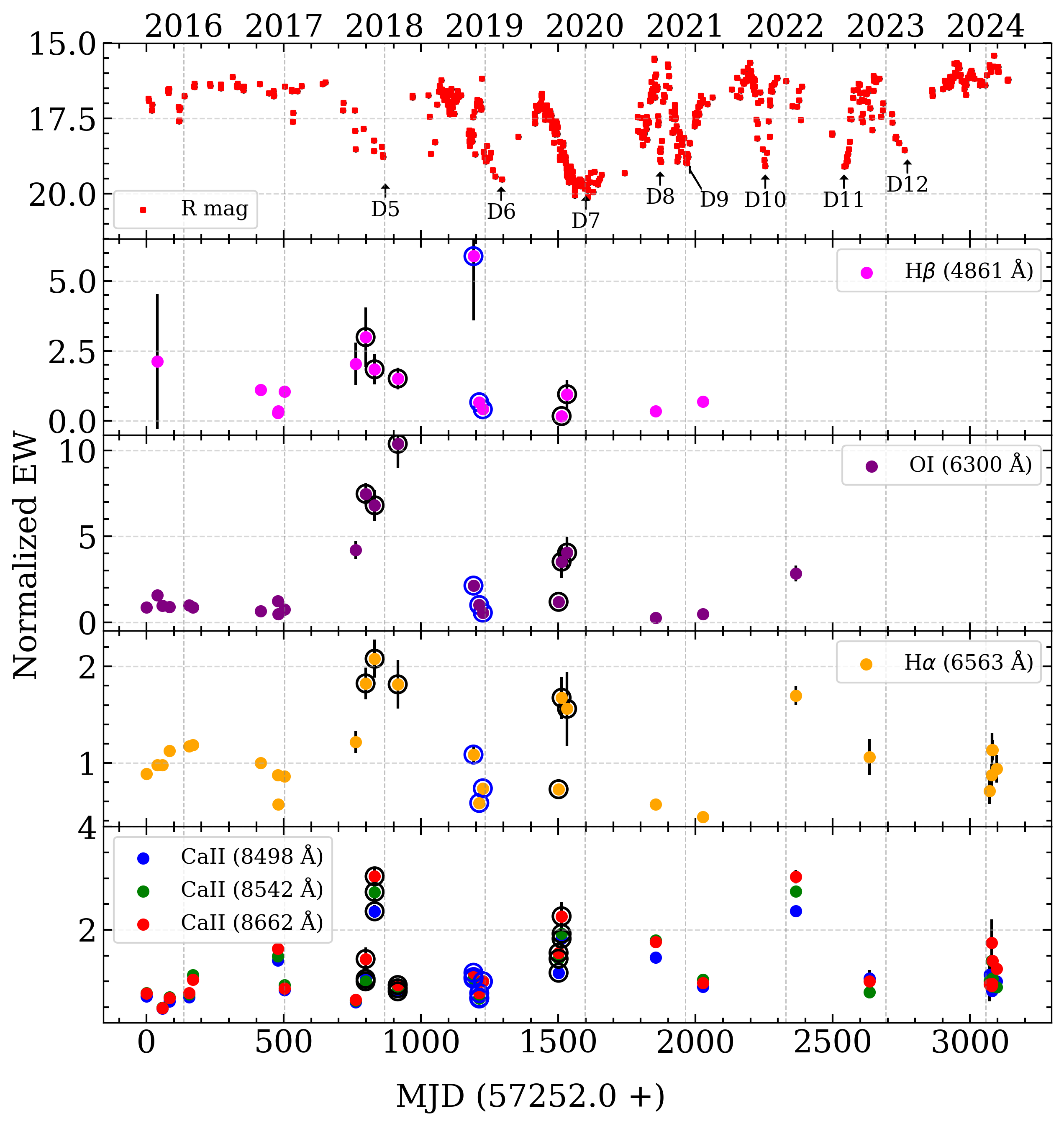}}
     \subfigure{\includegraphics[width=0.49\linewidth]{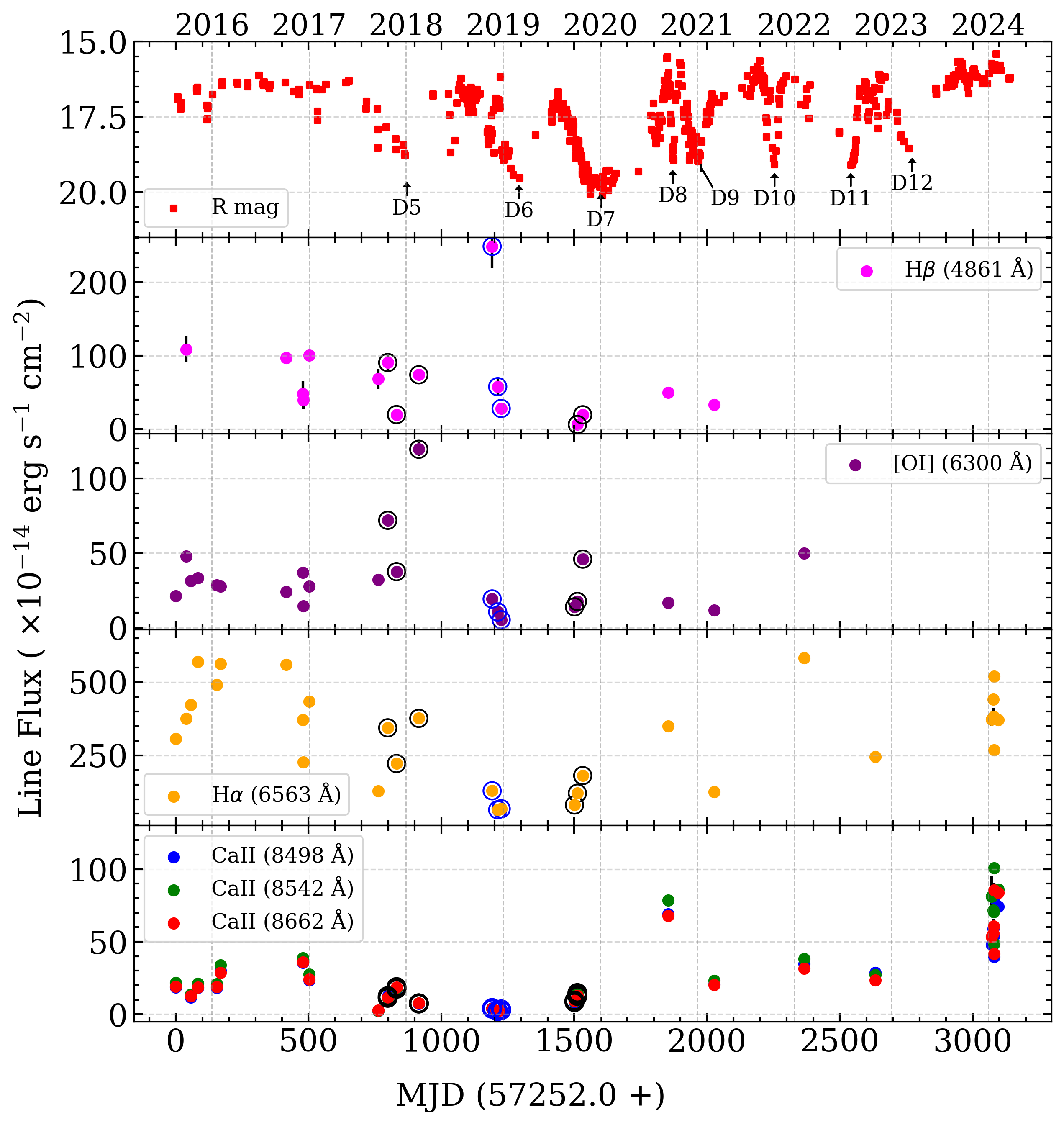}}
         \subfigure{\includegraphics[width=0.49\linewidth]{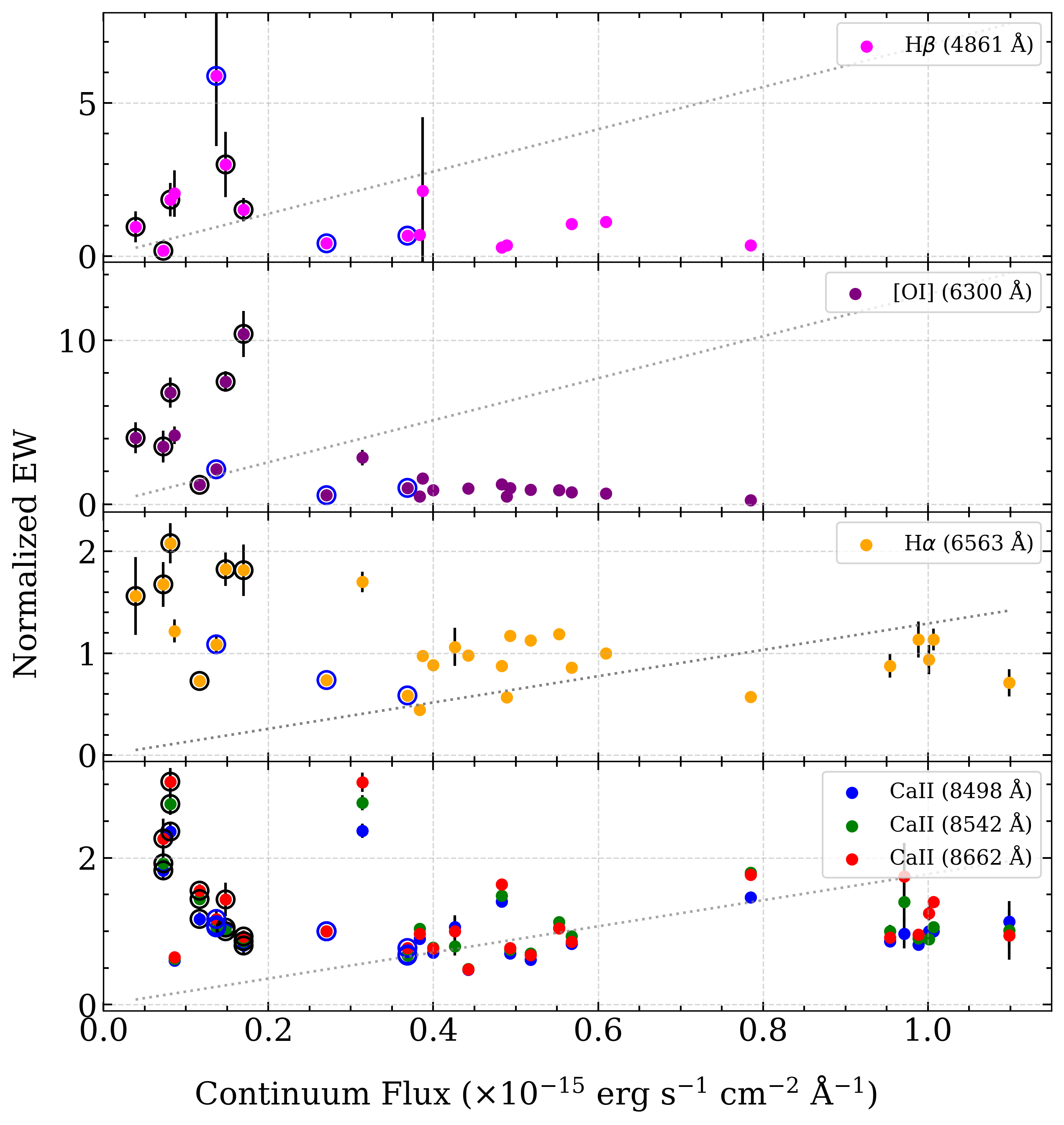}}
     \subfigure{\includegraphics[width=0.49\linewidth]{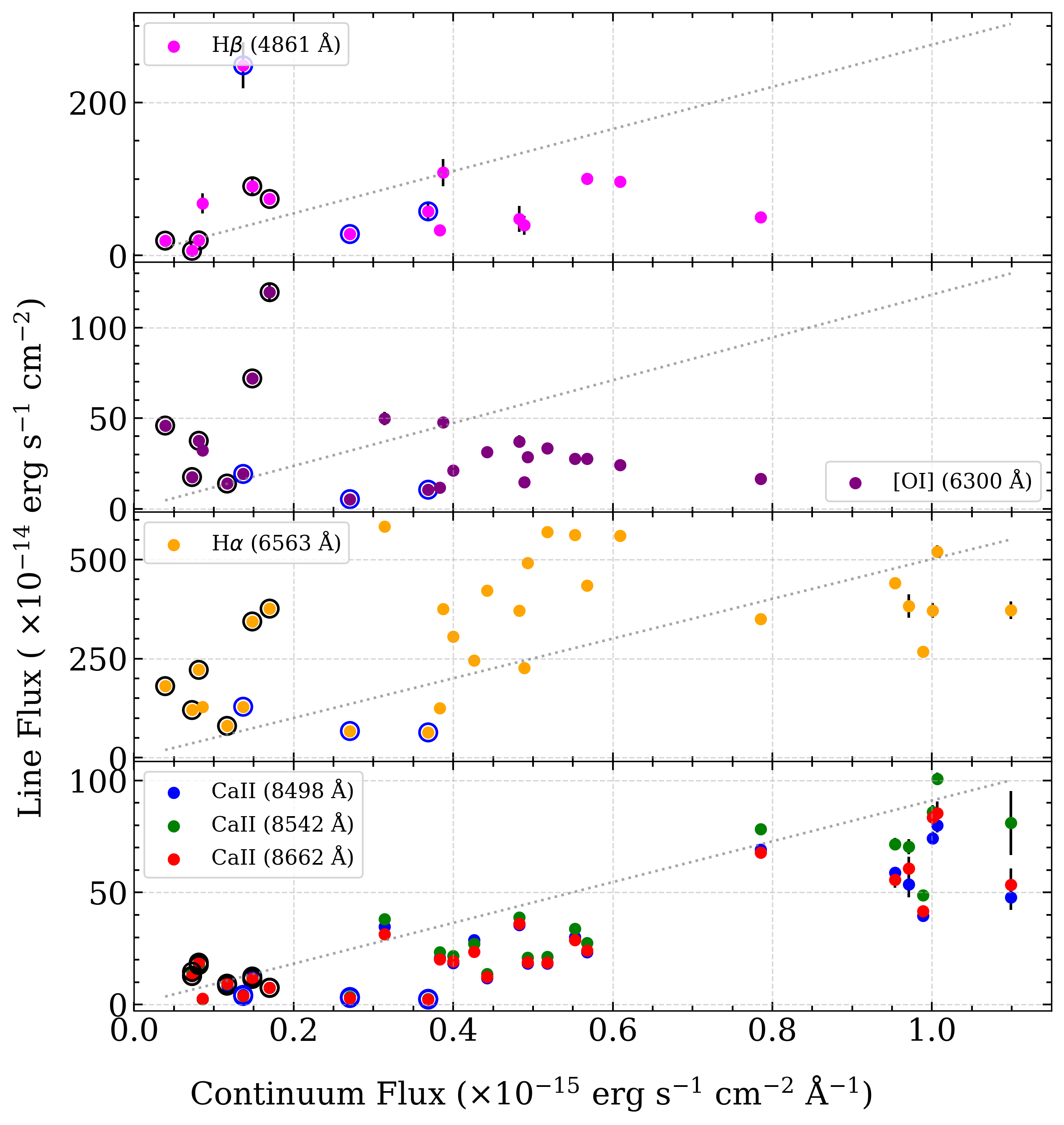}}
    \caption{\textit{Left panels}: The median normalized EWs of the H${\beta}$ (4861 \r{A}), [O I] (6300 \r{A}), H${\alpha}$ (6563 \r{A}), and Ca II IR triplet lines are shown as a function of MJD (upper panel), and flux derived from the R-band magnitude (combined from \citealt{Mutafov_2022RAA....22l5014M} and ZTF) in lower panel. \textit{Right panels}: The line flux of the same lines is plotted against MJD and flux. In the upper left and upper right panels, the LC using R-band mag (combined from \citealt{Mutafov_2022RAA....22l5014M} and ZTF) is shown to compare the variations in EW and line flux with photometric changes. The dips identified are also marked in these sub-panels. In the lower panels, the grey dotted lines represent lines with a slope of 1, indicating equal relative changes along both axes. EWs and line flux during dips D5 and D7 are marked with black circles, while those during D6 are marked with blue circles. (The calculated line flux and EW data are available.)
    }
    \label{fig:ew_mjd}
\end{figure*}

\subsubsection{Equivalent Widths and Line Fluxes}

To investigate the variability of prominent spectral features, we measured both the line fluxes and EWs of key emission lines. While the line fluxes quantify the intrinsic variability of the emission regions, the comparison between variations in EWs and fluxes provides insights into the underlying variability mechanism---such as distinguishing between variable accretion and extinction---and whether the occulting material is obscuring the continuum source, the line-forming region, or both.
The EWs were calculated using PHEW\footnote{\url{https://github.com/CoolStarsCU/PHEW}} \citep{Alam_2016zndo.....47889A}, which fits Gaussian profiles to emission lines and estimates EWs and uncertainties via Monte Carlo iterations. Spectral cutouts were manually defined by selecting wavelength windows around each line, with local continua estimated from regions of 10 \r{A} on either side. Line fluxes were computed using custom Python scripts that adopt similar techniques to PHEW. For the [S II] doublet (6717, 6731 \r{A}), we employed a two-Gaussian component fit to accurately model the line profile. All line fluxes were corrected for extinction using the $A_V$ value corresponding to each observing epoch (see Section \ref{subsec:cc}). The derived EWs and line fluxes are presented in Figure \ref{fig:ew_mjd} and available electronically as data behind the figure.

The EW of the H$\alpha$ line spans from $\sim$ -92 \r{A} to  -586 \r{A}, with a median of  -250 \r{A}. These values are consistent with those reported by \citet{Antoniucci_2014A&A...565L...7A}, who measured $\sim$ -400 \r{A} in a bright state ($I\sim$15.2 mag). \citet{Kun_2011} reported a maximum of  -900 \r{A} in 2008, also during a bright phase. In our observations, the highest H$\alpha$ EWs occur during low-brightness states (D5 and D7), suggesting that these increases are primarily due to reduced continuum levels from extinction, rather than intrinsic changes in line emission. The lowest EWs are observed in the post-D9 brightening phase, indicative of continuum recovery.
The Ca II IR triplet shows EW values ranging from  -15 \r{A} to  -90 \r{A}, with a median around  -35 \r{A}. Higher EWs are seen during major dimming episodes (D5 and D7), while lower values are found during intermediate-brightness phases (e.g., $ R\sim$17.5 mag in October 2015), during a shallow dip just before D5 (MJD 58014), and during D6. Notably, during the brightening near MJD 59106 (pre-D8), the Ca II EW reaches an intermediate value of  -50 \r{A}, implying contributions from both extinction changes and accretion variability.
The [O I] 6300 \r{A} line EW generally mirrors that of H$\alpha$, peaking during faint states and supporting the idea that its emission is less affected by extinction compared to the stellar continuum.

Despite the limited temporal coverage in the NIR, with most observations from January-February 2024 and one from November 2022, variability in Paschen lines is evident. EWs for the Paschen series are typically below  -20 \r{A}, with Pa5 reaching  -19 \r{A}. Br$\gamma$ shows much less variability, maintaining a near-constant EW around -5 \r{A}, suggesting a more stable emission region.

Figure~\ref{fig:ew_mjd} shows EW and line flux variations for H$\beta$, [O I], H$\alpha$, and the Ca II IR triplet over time, alongside the R-band LC from \citet{Mutafov_2022RAA....22l5014M} and ZTF data.
During D5 (Dec 2017--Jan 2018), a sharp increase in EWs of [O I], H$\alpha$, and Ca II is observed relative to 2016. Such increases can arise from (1) enhanced emission, or (2) reduced continuum due to circumstellar obscuration, as per the puffed-up inner disk model of \citet{Dullemond_2003ApJ...594L..47D}, commonly invoked in UXor-type stars \citep[e.g.,][]{Giannini_2016A&A...588A..20G}. In this case, line fluxes remain similar to 2016, implying the EW increase is driven by continuum suppression. A similar behavior is noted in D7 (highlighted by black circles in Figure~\ref{fig:ew_mjd}).

\begin{figure*}
    \centering
    \includegraphics[width=\linewidth]{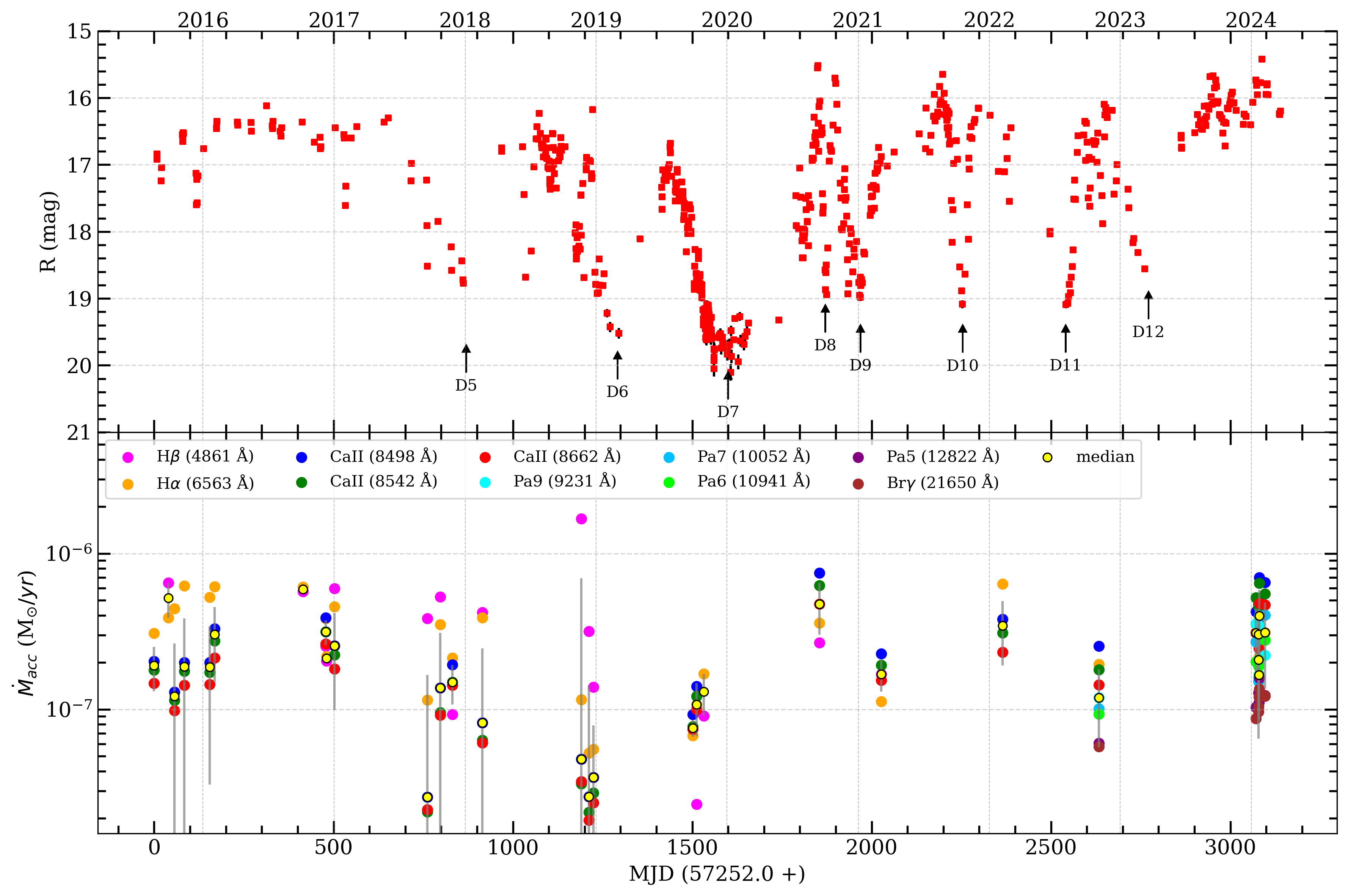}
    \caption{\textit{Lower panel:} The accretion rates estimated from various lines from the HFOSC and TANSPEC spectra of V1180 Cas are plotted against MJD. Accretion rates are plotted for different epochs. The yellow points are the median of accretion values in each epoch. The standard deviation in accretion values within each epoch is shown as error bars using solid vertical lines in grey color. \textit{Upper panel:} The LC using R-band magnitude (combined from \citealt{Mutafov_2022RAA....22l5014M} and ZTF) is shown to compare the variation in the accretion rate with photometric changes. The dips identified in the LC are also marked in this panel. (The calculated accretion rate data are available.) }
    \label{fig:accrtn_all}
\end{figure*}

D6 (Dec 2018--Jan 2019), though photometrically similar to D5, shows significantly different spectroscopic features. Ca II EWs are similar to or lower than those from 2016, and line fluxes are an order of magnitude lower than 2016 values. The $\Delta$R$\sim$2.5 mag drop suggests both extinction and a decline in accretion contribution, supported by the drop in accretion rate (Figure~\ref{fig:accrtn_all}).

A notable brightening event between 2020 and 2021 precedes D8, also reported by \citet{Mutafov_2022RAA....22l5014M}. This phase shows increased Ca II line fluxes and strong higher-order Paschen lines (Pa12, Pa11, Pa10), indicating a spike in accretion. This brightening is likely due to both enhanced accretion and a temporary clearing of obscuring material.

One interesting aspect of [O I] 6300 \r{A} is its behavior during brightness dips. In UXors like RR Tau \citep{Grinin_2023MNRAS.524.4047G}, increased EWs during dips coincide with constant line fluxes, suggesting extinction without a change in line emission. In V1180 Cas, however, both EWs and line fluxes of [O I] increase during D5 and D7, implying enhanced emission possibly linked to the obscuring material itself (e.g., outflowing gas/dust). Alternatively, partial coverage of the [O I]-emitting region may complicate the extinction behavior, deviating from standard UXor scenarios.

\begin{figure}
    \centering
    \includegraphics[width=\linewidth]{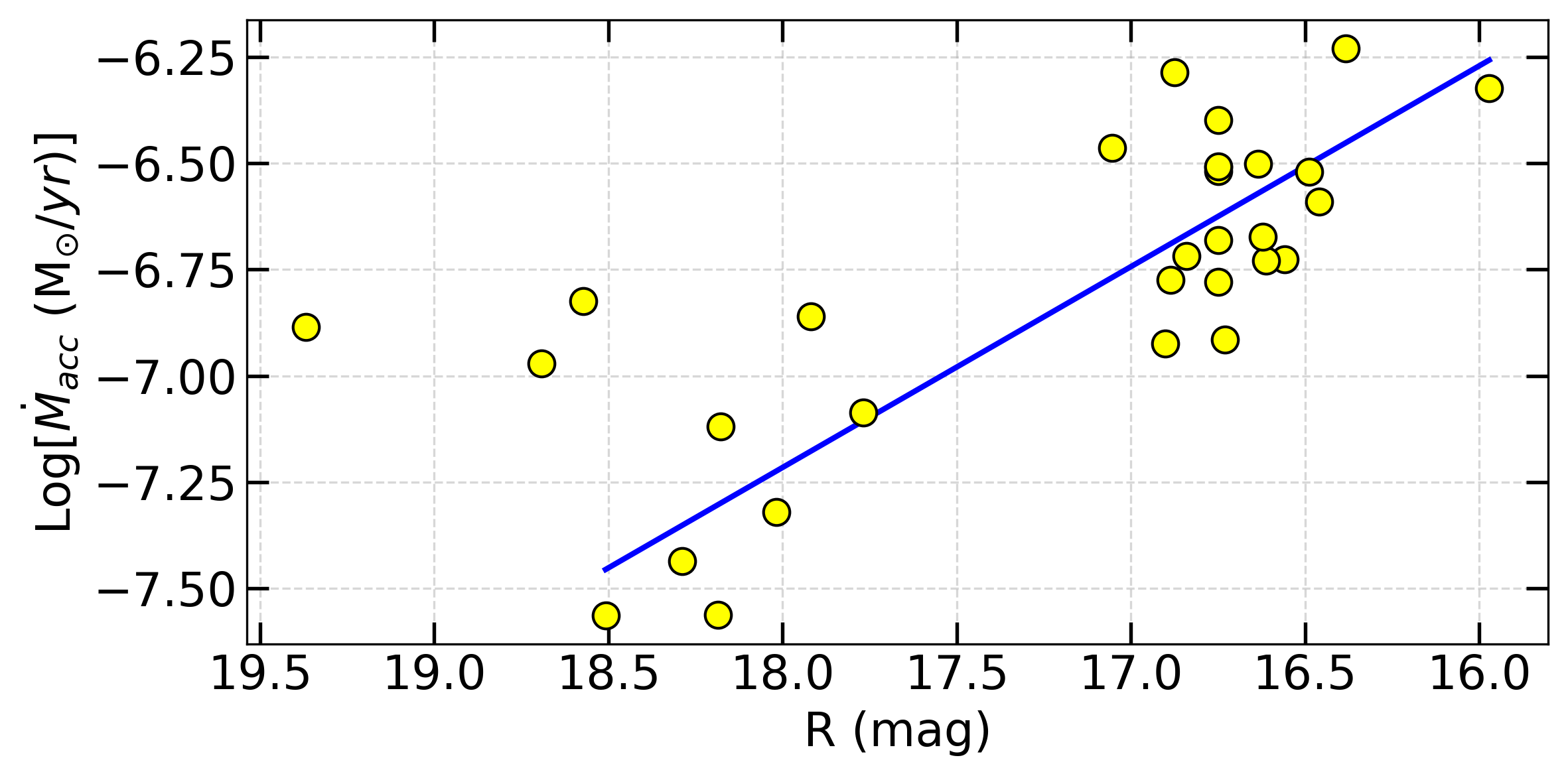}
    \caption{The median of accretion values from various lines in each epoch using the HFOSC and TANSPEC spectra of V1180 Cas, plotted against R-band magnitude (combined from \citealt{Mutafov_2022RAA....22l5014M} and ZTF). The solid blue line shows the linear fit (slope = $-0.47 \pm 0.05$ $M_\odot$yr$^{-1}$mag$^{-1}$) of the data points (excluding three data points which are fainter, $R >18.5$ mag and have accretion rate greater than $10^{-7.5}$$M_{\odot}$yr$^{-1}$. }
    \label{fig:accrtn_mag}
\end{figure}

\subsubsection{Accretion Rates}

To estimate the mass accretion rate, we first calculated the accretion luminosity using empirical relations between line luminosity (L$_{\text{line}}$) and accretion luminosity (L$_{\text{acc}}$), as derived by \cite{Alcala_2014A&A...561A...2A}:

\begin{equation} 
\log\left(\frac{L_{\text{acc}}}{L_{\odot}}\right) = a \times \log\left(\frac{L_{\text{line}}}{L_{\odot}}\right) + b 
\end{equation}

The coefficients $a$ and $b$ were adopted from \cite{Alcala_2017A&A...600A..20A}. To compute L$_{line}$, we first de-reddened the observed line flux using time-varying A$_{V}$ values. We derived the extinction values by first computing the $\Delta$A$_{V}$ for each epoch (see Section \ref{subsec:cc}), taking the mode of these differences (0.88 mag) as corresponding to A$_{V}$ = 4.6 mag \citet{Kun_2016ApJ...822...79K}. A$_{V}$ values for other epochs were then scaled accordingly and linearly interpolated to match the epochs of the spectroscopic observations.
The de-reddened line flux was then converted to line luminosity by scaling it with a factor of $4\pi d_{\ast}^{2}$, where $d_{\ast}$ is the distance to V1180 Cas. We adopted $d_{\ast}$ = 825 pc \citep{Kun_2016ApJ...822...79K}. Finally, the mass accretion rate was calculated using the relation:

\begin{equation} 
\dot{M}_{\text{acc}} = \frac{L_{\text{acc}} R_{\ast}}{G M_{\ast} \left(1 - \frac{R_{\ast}}{R_{\text{in}}}\right)} 
\end{equation}

For the stellar radius (R$_{\ast}$) and inner disk radius (R$_{\text{in}}$) of V1180 Cas, we adopted the values used by \citet{Antoniucci_2014A&A...565L...7A} and \citet{Kun_2011}, i.e., $R_{\ast}$ = 2 $R_{\odot}$ and R$_{\text{in}}$ = 5 $R_{\ast}$. The stellar mass was taken as the revised value from \citet{Kun_2016ApJ...822...79K}, i.e., $M_{\ast}$ = 0.65 $M_{\odot}$.

The accretion rates derived from different lines across various epochs using HFOSC and TANSPEC spectra are illustrated in Figure \ref{fig:accrtn_all} and available electronically as data behind the figure.
 
The lower panel of Figure \ref{fig:accrtn_all} shows calculated accretion rates from various spectral lines, spanning optical to NIR wavelengths, over several epochs. The yellow data points in this panel represent median accretion values for each epoch, with standard deviation as grey error bars. The upper panel in this figure shows the R-band LC to compare photometric change with the change in accretion.
In the lower panel of Figure~\ref{fig:accrtn_all}, a clear decline in accretion rates is observed during the dips in the LC.

The median values of accretion rate calculated from different lines on various epochs vary from $\sim$$2.7 \times 10^{-8}$ $M_{\odot}$yr$^{-1}$ up to $\sim$$5.8 \times 10^{-7}$ $M_{\odot}$yr$^{-1}$ with a median value of $\sim$$1.9 \times 10^{-7}$ $M_{\odot}$yr$^{-1}$.

\cite{Kun_2011} reported accretion rates during the bright state ($\sim$15 mag in $I$ band) in 2008 and 2009 to be approximately $1.6 \times 10^{-7}$ $M_{\odot}$yr$^{-1}$. In contrast, during September 2005, when the $I$-band magnitude was around 18 mag, the accretion rate was estimated to be about ten times lower. Based on this, they concluded that the majority of the observed photometric variability is driven by changes in the accretion rate. To account for the full amplitude of brightness variation ($\Delta I \sim 5$ mag), they suggested that the accretion rate must vary by a factor of 100. However, subsequent observations by \citet{Antoniucci_2014A&A...565L...7A}, obtained during a bright state in late 2013 using NIR spectroscopy, reported accretion rates that were an order of magnitude lower than those estimated by \citet{Kun_2011} from optical photometry and spectroscopy during earlier active phases. This discrepancy could arise from differences in the observational epochs, indicating intrinsic variability in the accretion process, or from the use of different diagnostic tracers and methodologies, which are sensitive to different regions and physical conditions within the accretion flow.

In our analysis, the accretion rates estimated during the bright phase of 2016–2017 are consistent with those reported by \citet{Kun_2011}, on the order of $10^{-7}$ $M_{\odot}$yr$^{-1}$. A significant drop in the accretion rate---by approximately one order of magnitude—is observed during the period spanning December 2018 to January 2019 (D6), coinciding with a photometric decline of $\Delta R \sim 2.5$ mag in the $R$ band. This suggests a strong correlation between the star's optical brightness and its accretion activity. Interestingly, during the earlier and deeper dimming episode between December 2017 and January 2018, the corresponding decrease in accretion rate appears less pronounced, indicating that other factors---such as variable extinction or viewing geometry---may also contribute to the observed photometric variability.

In Figure~\ref{fig:accrtn_mag}, the median accretion rates for each observing epoch are plotted against the corresponding $R$-band magnitudes. A linear fit to the logarithmic accretion rate versus $R$-band magnitude yields a slope of approximately  $-0.47 \pm 0.05$ $M_\odot$yr$^{-1}$mag$^{-1}$(blue solid line). The negative slope indicates a strong correlation (with Pearson's correlation coefficient r$\sim-0.9$), suggesting that variations in the accretion rate significantly contribute to the observed changes in the optical brightness of V1180 Cas. However, the presence of a few outliers deviating from this trend implies that additional factors---such as variable circumstellar extinction or viewing geometry---may also play a role in modulating the observed flux.

\subsubsection{Outflow rate, Density, and Temperature}

We estimate the mass-loss rate using the [O I] (6300 \r{A}) emission line flux, applying the method described in relation A8 from \cite{Hartigan_1995ApJ...452..736H}. Since optically thin forbidden line emissions that originate in outflows are directly linked to mass ejection, this flux serves as a reliable tracer. Using HFOSC spectra obtained through a 1.95$^{\prime \prime}$ slit, we calculated a median mass outflow rate of  $\sim$1.2 $\times 10^{-8}$ $M_{\odot}$yr$^{-1}$. This estimate assumes a distance of 825 pc to the source and a projected outflow velocity component in the plane of the sky of 150 km s$^{-1}$ \citep{Hartigan_1995ApJ...452..736H, Ninan_2015ApJ...815....4N} from the central star.
The calculated outflow rate is around one order of magnitude lower than the accretion rate (i.e., $\sim$$1.9\times 10^{-7}$ $M_{\odot}$yr$^{-1}$). For Class II sources with accretion rate ranges from $\sim$$10^{-6} - 10^{-12}$ $M_{\odot}$yr$^{-1}$, the ratio of outflow rate to accretion rate for the high-velocity outflow component (jet) is found to be around 0.1. While for the low-velocity component (wind), it ranges from $\sim$$0.1-1$ (see Table 1 in \citealt{Pascucci_2023ASPC..534..567P} and references therein). 

Optically thin forbidden emission lines arise from the outflow and can be used to estimate the electron density and temperature of the emitting gas \citep{Ninan_2015ApJ...815....4N, Fang_2018ApJ...868...28F}.
The flux ratio of [S II] 6717 \r{A} to 6731 \r{A} serves as an effective diagnostic of electron density and is relatively insensitive to temperature variations. In our HFOSC spectra, we observed both emission lines prominently across multiple epochs, although they appear blended. To separate their individual contributions, we applied a two-component Gaussian fitting technique. The median flux ratio obtained is 0.62, which corresponds to an electron density of approximately 4 $\times 10^{3}$ electron cm$^{-3}$ in the shocked regions of the outflow \citep{Osterbrock_2006agna.book.....O}. Although the [O I] 5577 \r{A} line is not detected in the HFOSC spectra of V1180 Cas, we estimate an upper limit on the [O I] 5577 \r{A}/6300 \r{A} flux ratio by deriving an upper limit to the line flux ($F_{up}$) following the method described in \cite{2021A&A...650A..43F}. For the [O I] 5577 \r{A} line, we first fitted and subtracted the local continuum within the window of 100 \r{A} centered at 5577 \r{A}, then measured the root mean square (rms) in the continuum-subtracted spectrum. The upper limit flux for the line is computed as:
\begin{equation}
F_{\rm up} = 3 \times {\rm rms} \times \left(\frac{\lambda_{\rm line}}{R}\right),
\end{equation}
where $\lambda_{line}$ is central wavelength (i.e. 5577 \r{A}) and $R$ is resolution of the instrument. The resulting median ratio of [O I] 5577 \r{A}/6300 \r{A} $<$ 0.07 suggests an electron density of $<$ 1.2$\times$10$^{6}$  electron cm$^{-3}$ for a gas temperature of $10^{4}$ K \citep{Natta_2014A&A...569A...5N}. Assuming the electron density value $1.2\times 10^{6}$ electron cm$^{-3}$, the line ratio [O I] 5577 \r{A}/(6300+6364) \r{A} $<$ 0.05 implies a temperature $<$9500 K \citep{Osterbrock_2006agna.book.....O}. We have utilized the PyNeb\footnote{\url{https://github.com/Morisset/PyNeb_devel/tree/master/docs}} \citep{pyneb_2015A&A...573A..42L} python module to calculate the density and temperature using line flux ratio. 
Using high-resolution spectra of T Tauri stars, \citet{Fang_2018ApJ...868...28F} found that the observed forbidden emission line ratios in the low-velocity component are consistent with thermal excitation in gas at temperatures of $5000-10000$ K and electron densities of approximately $10^{7}-10^{8}$ electron cm$^{-3}$. They further demonstrated that the line ratios in the high-velocity component can be explained by radiative shock models, where the emission arises from hot $\sim$$7000-10000$ K, but less dense, post-shock cooling regions. 

\section{Discussion}
\label{sec:discussion}
\subsection{Photometric properties}

The long-term photometric evolution of V1180 Cas from 1999 to 2025 reveals a dynamic and complex variability pattern, reflecting significant changes in its circumstellar environment. Initially, between 1999 and ~2011, the LC showed three major optical dips (D1–D3) in the I band, each with decreasing duration and depth. These early dimming events, reported by \citet{Kun_2011}, are consistent with episodic increases in circumstellar extinction. Subsequent observations by \citet{Mutafov_2022RAA....22l5014M} revealed further dips (D4–D7), suggesting that the variability persisted in the following decade.

From 2018 onward, the source entered a more active phase characterized by eleven prominent dimming events (D5–D15), occurring quasi-periodically on timescales of $\sim$1 year or shorter. These dips, captured with high temporal cadence by surveys such as ZTF, Gaia DR3, ATLAS, and ASAS-SN, are often asymmetric—showing slower fading and faster recovery—and are accompanied by stochastic variability of up to $\sim$1.5 mag on short timescales (days to weeks). Such patterns are typical of UXor-type YSOs, where circumstellar dust structures intermittently obscure the central star.

The brightening and dimming rates observed in V1180 Cas are broadly comparable to those reported for other EXor- and UXor-type YSOs (e.g., EX Lup, Gaia 20eae, V2492 Cyg, GM Cep; \citealt{Hillenbrand_2013AJ....145...59H, Abraham_2018ApJ...853...28A, Huang_2019ApJ...871..183H, Ghosh_2022ApJ...926...68G, Miera_2023A&A...678A..88C}). Typical rates in these systems range from 0.01 to 0.1 mag day$^{-1}$, consistent with the moderate variations seen in V1180 Cas, although in certain episodes (e.g., MJD 59100–59170), the rates exceed 0.2 mag day$^{-1}$, suggesting more dynamic events, possibly indicative of enhanced accretion bursts or rapid disk reconfigurations.

Linear fits to the fading and brightening phases of these dips reveal a range of variability rates, from as low as 0.01 mag day$^{-1}$ to over 0.23 mag day$^{-1}$ during rapid transitions (e.g., D8). Notably, there appears to be an evolution in these rates over time: earlier dips exhibited faster fading and slower brightening, whereas more recent dips show the reverse. This temporal trend may indicate a possible change in dominant variability mechanism, suggesting a change in the disk structure, the distribution of inner disk material, or fluctuations in accretion activity. Alternatively, variations in the sharpness of cloud edges or in their orbital velocities could also account for the observed evolution in the fading and brightening rates.

A particularly intriguing feature is the detection of a 29.2-day periodic signal in the residual LC (58665-58820 MJD) during dip D7. This may reflect the rotational modulation of disk structures or the orbital motion of circumstellar clumps close to the sublimation radius. However, the absence of a persistent periodicity across the full dataset suggests that the observed variability is governed by a combination of transient, non-periodic events superimposed on a longer-term, evolving baseline. It is important to note that this interpretation holds only if the observed periodicity is intrinsic to the system and not an artifact of lunar phase variations or other observational systematics.

The CMDs and CC diagrams in the optical, NIR, and MIR regimes provide important constraints on the nature of the variability. In the optical, V1180 Cas generally reddens as it fades—consistent with increasing extinction—but also shows blueing episodes during several dips (e.g., late 2019, early 2021, August 2023). These color reversals, observed in the ZTF CMD, are classic signatures of UXor-type behavior, attributed to the increased dominance of scattered light when the direct stellar flux is partially blocked by circumstellar material. Similar behavior has been reported in V1184 Tau, GM Cep, and V852 Aur. These phenomena indicate a clumpy and evolving inner disk, where optically thick dust structures intermittently obscure the star, while scattering halos continue to contribute to the observed flux.

In contrast, the MIR and NIR colors consistently show reddening during fading events, with no blueing trend. This is expected, as MIR/NIR wavelengths are less sensitive to scattering and more directly trace changes in thermal emission from the disk. The NEOWISE MIR LC (W1 band), color-coded by W1–W2, shows a strong correspondence with the optical LC, 
implying that both regimes respond simultaneously to changes in the disk. Interestingly, a long-term linear brightening trend is seen in the MIR between 2017 and 2023, possibly indicating a gradual increase in inner disk emission, either due to enhanced heating or structural changes in the disk's inner rim.

Quantitatively, the extinction changes required to reproduce the observed brightness variations differ substantially across wavelengths---$\Delta A_V \sim 3$ mag in the optical, $\sim$10 mag in the NIR, and $\sim$30 mag in the MIR. Using standard extinction law, $\Delta A_V \sim 3$ mag corresponds to $\Delta J$ = 0.73 mag, $\Delta H$ = 0.39 mag, $\Delta K$ = 0.23 mag, $\Delta W1$ = 0.12 mag, and $\Delta W2$ = 0.08 mag. These predicted MIR amplitudes are much smaller than the observed variations ($\Delta W1 \sim$1 mag and  $\Delta W2 \sim$0.5), demonstrating that extinction alone cannot explain the MIR variability. Instead, the observed changes across optical–MIR wavelengths are likely driven by variations in the effective temperature and structure of the emitting regions, possibly linked to transient heating or reconfiguration of the inner disk.

The NIR CMD and CC diagrams further reinforce this picture. As the source fades (J $>$ 15.5 mag), it moves away from the classical T Tauri locus toward redder colors, accompanied by signs of IR excess, consistent with variable extinction and changes in the disk's thermal structure. Such motion in the color space cannot be explained by extinction alone and suggests that the temperature or geometry of the inner disk is evolving, possibly due to localized accretion bursts or reorganization of material at the disk's inner edge.

Taken together, these multi-wavelength observations highlight V1180 Cas as a hybrid object exhibiting both UXor- and EXor-like characteristics. The quasi-periodic dips and blueing behavior are hallmarks of UXors dominated by variable extinction from disk warps or clumps. Meanwhile, the presence of brightening episodes, stochastic variability, and rapid changes in the MIR LC point to episodic accretion events more typical of EXors. Such hybrid behavior suggests that both disk obscuration and unsteady accretion processes play significant and interconnected roles in driving the observed variability of V1180 Cas.
 
\subsection{Spectroscopic Variability of V1180 Cas: Interplay of Accretion, Outflow, and Extinction}

The rich spectroscopic dataset for V1180 Cas, spanning optical to NIR (0.5–2.5 $\mu$m) wavelengths over more than 30 epochs, reveals a complex interplay between accretion, outflow, and circumstellar extinction processes. These observations, acquired over nearly a decade and during various photometric states, provide valuable insights into the dynamic environment and physical conditions of this highly variable YSO.

A key finding is the persistent presence of strong emission lines at all epochs, including hydrogen recombination lines (H$\alpha$, Pa$\beta$, Br$\gamma$), the Ca II IR triplet, He I (1.08 $\mu$m), and various forbidden lines such as [O I] 6300 \r{A}, [S II] 6717, 6731 \r{A}, and [Fe II] 1.64 $\mu$m. The hydrogen lines, primarily tracing magnetospheric accretion, exhibit line luminosities that correlate with changes in optical brightness. During bright phases, enhanced line fluxes and the appearance of higher-order Paschen lines suggest elevated accretion rates, typically a few times $10^{-7}$$M_\odot$yr$^{-1}$. In contrast, fainter states show a decline in accretion indicators with rates dropping to $\sim 10^{-8}$ $M_{\odot}$ yr$^{-1}$, although in some cases the EWs of emission lines increase due to a suppressed continuum, underscoring the need to disentangle intrinsic accretion variability from extinction-driven effects.

A particularly notable result is the increase in EWs of lines like H$\alpha$ and Ca II during major dimming episodes (e.g., D5 and D7) without a corresponding rise in line flux—indicative of extinction-dominated variability. Conversely, during D6 (Dec 2018–Jan 2019), both the EWs and line fluxes of Ca II decreased, pointing to a genuine reduction in accretion activity. A brightening episode in 2020–2021, preceding D8, is characterized by increased Ca II line fluxes and strong Paschen line emission, consistent with a significant accretion outburst.

The He I 1.08 $\mu$m line often displays a P-Cygni profile, signaling outflowing material along the line of sight. Interestingly, this absorption feature is not seen in H$\alpha$, implying that the H$\alpha$ arises from the more closer region to the magnetosphere while He I traces a different wind component—possibly arising from the outer regions.
Forbidden lines, particularly [O I] 6300 \r{A}, are detected at all epochs, including during deep dips in the LC. Unlike classical UXor-type-stars---where forbidden-line fluxes remain relatively constant and only their EWs increase during extinction events---V1180 Cas exhibits simultaneous increases in both EW and line flux during dips (e.g., D5 and D7). This suggests that the obscuring material may be physically associated with the outflow, or that dynamical variations in the inner disk simultaneously increase both extinction and outflow activity. Such behavior departs from standard UXor models and implies a more intricate coupling between disk winds and inner disk structures.

Additional insights are provided by the detection of CO overtone bandhead emission in the NIR spectra, indicative of hot molecular gas in the inner disk. Although all available TANSPEC spectra are obtained during relatively bright states, the presence of strong CO emission during these epochs suggests thermal excitation driven by enhanced stellar irradiation and accretion luminosity. Weak H$_2$ emission is also occasionally detected, pointing to a warm molecular component likely located farther out in the disk or from jet/outflow activity as previously observed by \cite{Antoniucci_2014A&A...565L...7A}.

The contrasting behaviors observed during different dimming events suggest that distinct physical mechanisms---circumstellar extinction and accretion variability---dominate at different times. Taken together, the multi-epoch variability supports a dual-mode scenario: one driven by variable extinction due to circumstellar dust (UXor-like), and the other by episodic enhancements in accretion activity (EXor-like).

The derived accretion rates---typically a few times  $10^{-7}$$M_\odot$yr$^{-1}$ are consistent with values observed in EXors \citep[e.g.,][]{Fischer_2023ASPC..534..355F}. For comparison, EX Lup has shown accretion rates of $\sim$9$\times$10$^{-7}$ $M_{\odot}$yr$^{-1}$ during its large 2008 outburst, $\sim$2$\times$10$^{-7}$ $M_{\odot}$yr$^{-1}$ during the medium 2022 burst, and $\sim$10$^{-8}$ $M_{\odot}$yr$^{-1}$ in quiescence \citep{2023A&A...678A..88C, Singh_2024ApJ...968...88S}.
During D6, the accretion rate dropped by an order of magnitude, concurrent with a 2.5 mag decrease in $R$-band brightness. The D5 (2017–2018) shows a less pronounced decline in accretion, reinforcing the role of variable extinction. 
A linear fit between accretion rate and $R$-band magnitude yields a slope of –0.47 $M_\odot$yr$^{-1}$mag$^{-1}$ , quantitatively supporting an accretion-dominated variability scenario with occasional deviations due to extinction or geometric effects.

The estimated mass outflow rate of $\sim$1.2$\times$$10^{-8}$$M_\odot$yr$^{-1}$ and an outflow-to-accretion ratio of $\sim$0.1 are consistent with magneto-centrifugally driven winds in classical T Tauri stars \citep[e.g.,][]{Pascucci_2023ASPC..534..567P}. This confirms a close coupling between accretion and outflow in V1180 Cas. Our derived outflow rate is somewhat higher but broadly consistent with values reported by \citet{Antoniucci_2014A&A...565L...7A}, where authors estimated mass-loss rates of $\sim$4$\times$10$^{-9}$ $M_\odot$yr$^{-1}$ from [O I] 6300 \r{A}, and [Fe II] 1.25 $\mu$m lines, and $\sim$4$\times$10$^{-10}$ $M_\odot$yr$^{-1}$ from H$_2$ 2.12 $\mu$m emission. The electron density of $\sim$$4\times10^3$ cm$^{-3}$, derived from the [S II] 6717 \r{A}/6731 \r{A} line ratio, is characteristic of moderately dense, shock-excited jets in low-mass YSOs. While the density and temperature estimated from [O I] lines suggest these lines trace thermally excited gas in the outflow.

In summary, V1180 Cas exhibits characteristics of both EXor and UXor classes, showing both accretion-driven and extinction-driven modes of variability at different epochs. Its spectral features, particularly the behavior of H I, Ca II, and forbidden lines, as well as the presence of molecular emission (CO and H$_2$), suggest a hybrid nature. Persistent outflow signatures and variable accretion rates further underscore the dynamic coupling between accretion, extinction, and mass-loss. Continued multi-wavelength, high-cadence monitoring is essential to disentangle these mechanisms and to understand whether the observed behavior represents episodic events or a longer-term evolutionary stage in disk-bearing YSOs.

\section{Summary and Conclusion} \label{sec:summ}

V1180 Cas was initially classified as an EXors-type source based on spectrophotometric data. However, subsequent analyses using long-term photometric monitoring and optical spectral energy distribution data led to its reclassification as a UXors-type source. 
We have conducted a decade-long follow-up of this source, reporting its long-term photometric and spectroscopic characteristics. Based on our analysis, we draw the following conclusions:

\begin{itemize}

\item 
V1180 Cas exhibits complex photometric behavior over two decades (1999-2025), marked by both long and short dimming events. Early dips (D1-D3) were sporadic and consistent with extinction, while recent dips (D5-D15) show quasi-periodicity and structured profiles.

\item
The variability is best explained by a dual-mode model, i.e.,  (a) UXor-like mode: Driven by circumstellar extinction due to dusty clumps or disk warps. (b) EXor-like mode: Dominated by episodic accretion bursts from the inner disk.

\item 
Optical CMDs show reddening during fades with blueing episodes--a hallmark of UXors. NIR and MIR colors consistently redden, indicating changes in thermal emission, not just extinction. A gradual MIR brightening trend suggests evolving disk structure or heating of the inner rim.

\item Strong emission lines (H I, Ca II, He I, [O I], [S II], [Fe II]) are persistently detected across 30 epochs. Hydrogen lines trace accretion, with luminosities that scale with brightness; Ca II and Paschen lines confirm variable accretion activity. Forbidden lines ([O I], [S II]) are enhanced during some dips, indicating dynamic linkages between outflow and extinction.

\item Equivalent widths and fluxes of emission lines evolve differently during different dips. D5/D7 show extinction-dominated behavior; D6 reflects a true drop in accretion rate. A linear correlation between accretion rate and $R$-band magnitude (slope $\sim$ -0.47 $M_\odot$yr$^{-1}$mag$^{-1}$) supports accretion-driven changes modulated by extinction.

\item He I P-Cygni profiles reveal inner disk winds. Estimated mass loss rate ($\sim$$1.2\times10^{-8}$ $M_\odot$yr$^{-1}$) and outflow-to-accretion ratio ($\sim$0.1) are consistent with magneto-centrifugal winds. Electron densities and the temperature estimate suggest [S II] traces moderately dense, shock-excited jets, while [O I] lines arise from thermally excited gas in the outflow. 

\item V1180 Cas exhibits a rare hybrid of UXor- and EXor-like behaviors, where extinction, accretion, and outflow processes appear to be tightly coupled and episodic in nature.

\end{itemize}

In conclusion, V1180 Cas appears to be an EXor-type eruptive star undergoing significant extinction, likely caused by circumstellar material, probably a puffed-up inner disk rim. Its variability timescales differ from those of typical EXors, possibly due to a combination of accretion-driven and extinction-driven processes. This behavior is reminiscent of V2492 Cyg, which exhibits characteristics of both EXors and UXors. To confirm the EXor nature of V1180 Cas, high-cadence, multi-wavelength photometric and spectroscopic monitoring---especially during phases of fading and brightening---is essential.

\section*{Acknowledgements}
We thank the anonymous reviewer for the comments and suggestions that enhanced the quality of the paper.
We thank the staff at the 1.3 m DFOT and 3.6 m DOT, Devasthal (ARIES), for their cooperation during observations. It is a pleasure to thank the members of 3.6 m DOT team and IR astronomy group at TIFR for their support during TANSPEC observations. TIFR-ARIES Near Infrared Spectrometer (TANSPEC) was built in collaboration with TIFR, ARIES, and MKIR, Hawaii, for the DOT. We thank the staff of IAO, Hanle, and CREST, Hosakote that made these observations possible. The facilities at IAO and CREST are operated by the Indian Institute of Astrophysics.
D.K.O., J.P.N., and K.S. acknowledge the support of the Department of Atomic Energy, Government of India, under project identification No. RTI 4002.
A.G. acknowledges support from DGAPA-UNAM and PAPIIT project IN105225.
RKY gratefully acknowledges the support from the Fundamental Fund of Thailand Science Research and Innovation (TSRI) (Confirmation No. FFB680072/0269) through the National Astronomical Research Institute of Thailand (Public Organization)
A.V. acknowledges the ﬁnancial support of DST-INSPIRE (No. DST/INSPIRE Fellowship/2019/IF190550).
M.C. acknowledges the financial support of DST-INSPIRE (No.: DST/INSPIRE Fellowship/2022/IF220587).

\textit{Facilities}: 1.3m DFOT, 3.6m DOT (TANSPEC), 2m HCT (HFOSC), ZTF, ATLAS, Gaia, 2MASS, NEOWISE, ASAS-SN.

\textit{Software}: Astropy \citep{Astropy_2013A&A...558A..33A}, Matplotlib \citep{Matplotlib_2007}, IRAF \citep{iraf1_1986SPIE..627..733T, iraf2_1993ASPC...52..173T}, DAOPHOT--II \citep{Stetson_1992ASPC...25..297S}.

\bibliography{references}{}
\bibliographystyle{aasjournal}

\appendix
\renewcommand\thefigure{A\arabic{figure}}
\setcounter{figure}{0}
\section{Short-term variability in WISE light curves}
\label{sec:appen_wise}
\begin{figure*}
    \centering
    \includegraphics[width=1\textwidth]{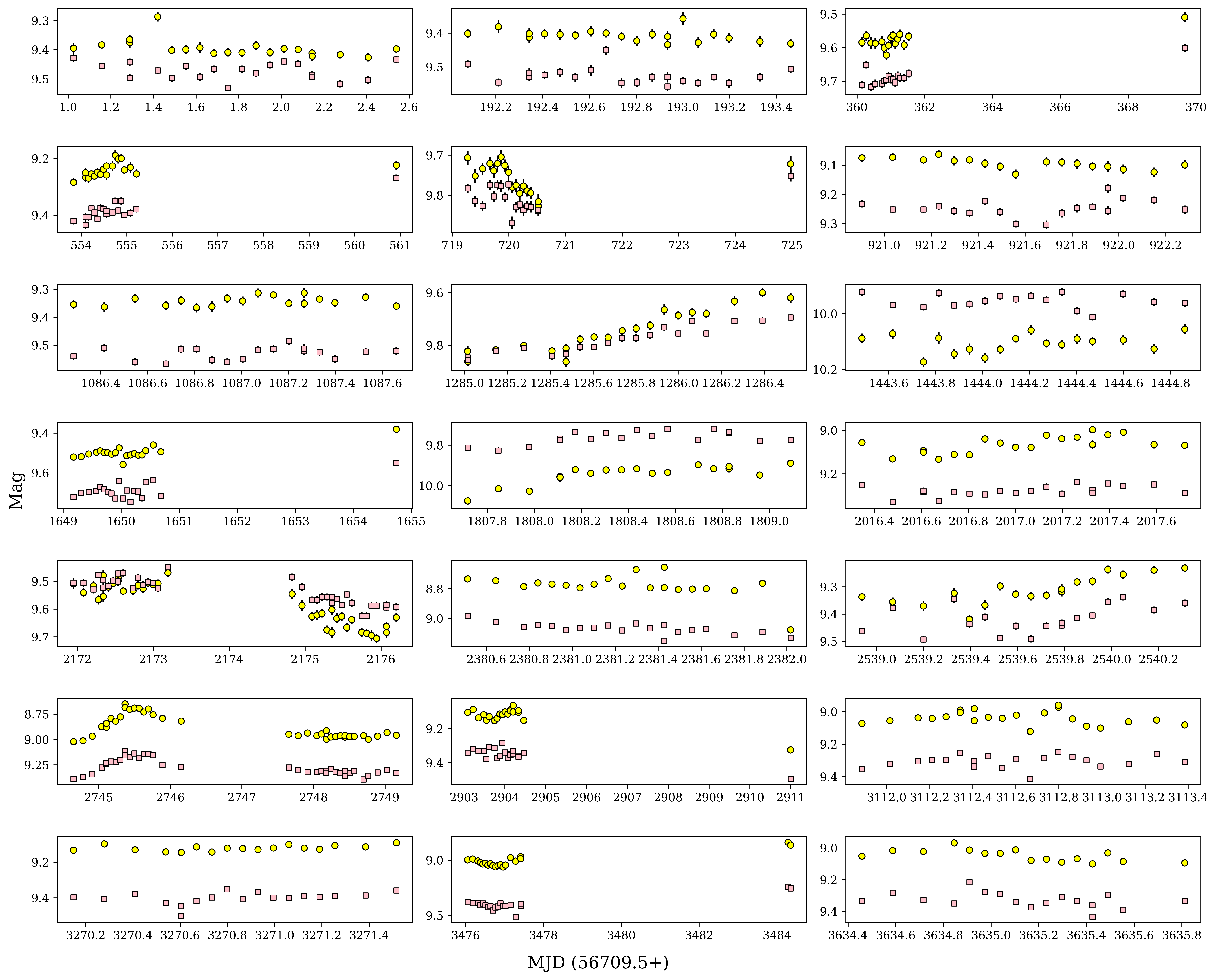}
    \caption{WISE ($W1$, $W2$) s of V1180 Cas. Yellow circles represent the $W1$-band data, while pink squares represent the $W2$-band data, shown with an offset of +1 mag for clarity. Each subplot displays data obtained within each observing window.}
    \label{fig:wise_lc}
\end{figure*}

V1180 Cas was observed with WISE in the $W1$ and $W2$ bands over 21 distinct observing windows between 2014 and 2024. Each window consists of multiple observations spanning 1–10 days, with successive windows separated by $\sim$150–200 days. These data are presented in Figure~\ref{fig:wise_lc} as individual subpanels. The WISE LC exhibits low-amplitude (a few mmag) variability on day timescales. In addition to this short-term variability, the LC occasionally displays linear trends. For instance, between MJD 57995 and 57996, the source brightened at a rate of $\sim$0.2 mag day$^{-1}$ in both bands. A similar event occurred near MJD 59454, with a brightening rate of $\sim$0.25 mag day$^{-1}$.

\begin{figure*}
    \centering
    \includegraphics[width=0.3\textwidth]{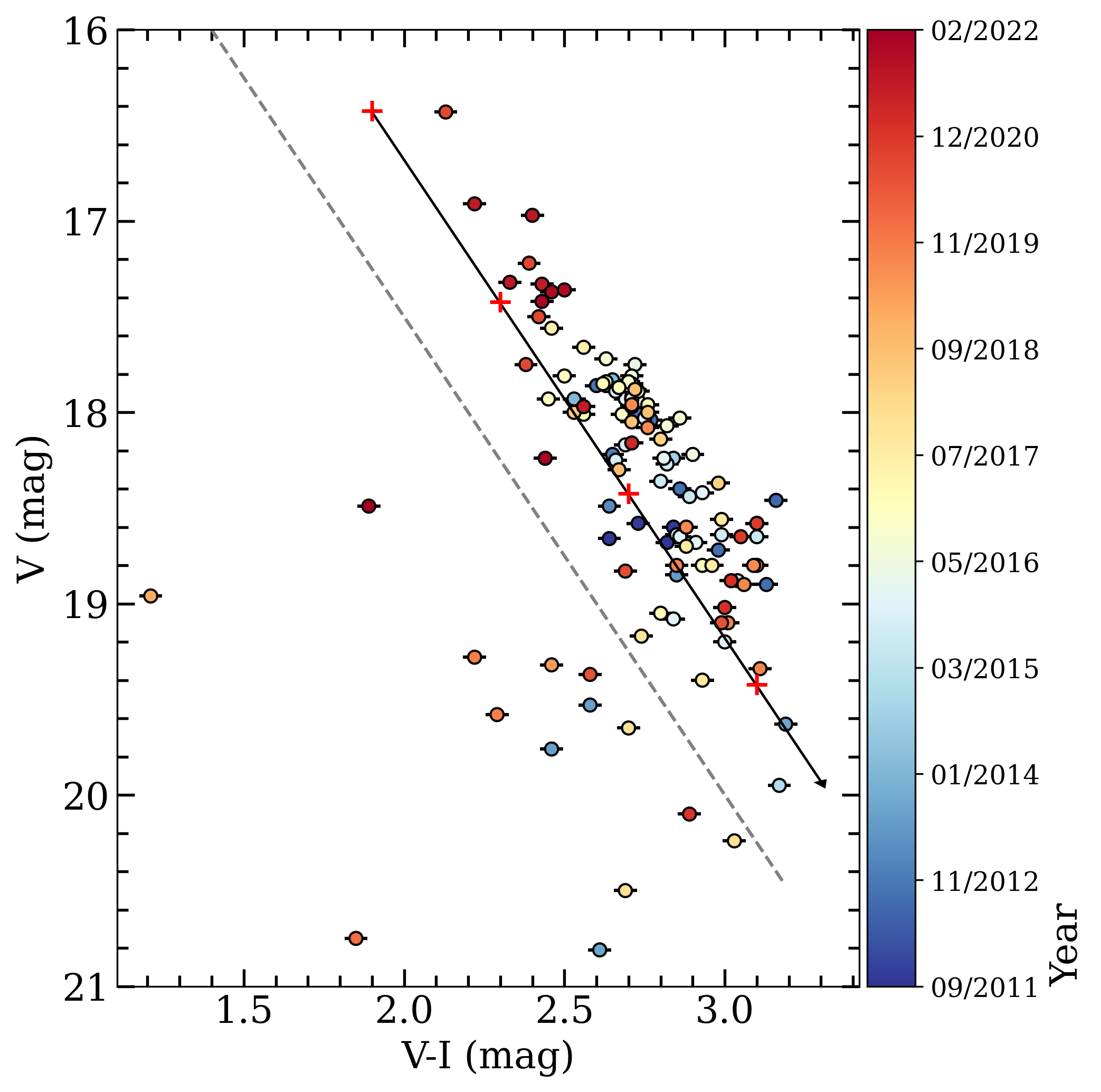}
    \includegraphics[width=0.3\textwidth]{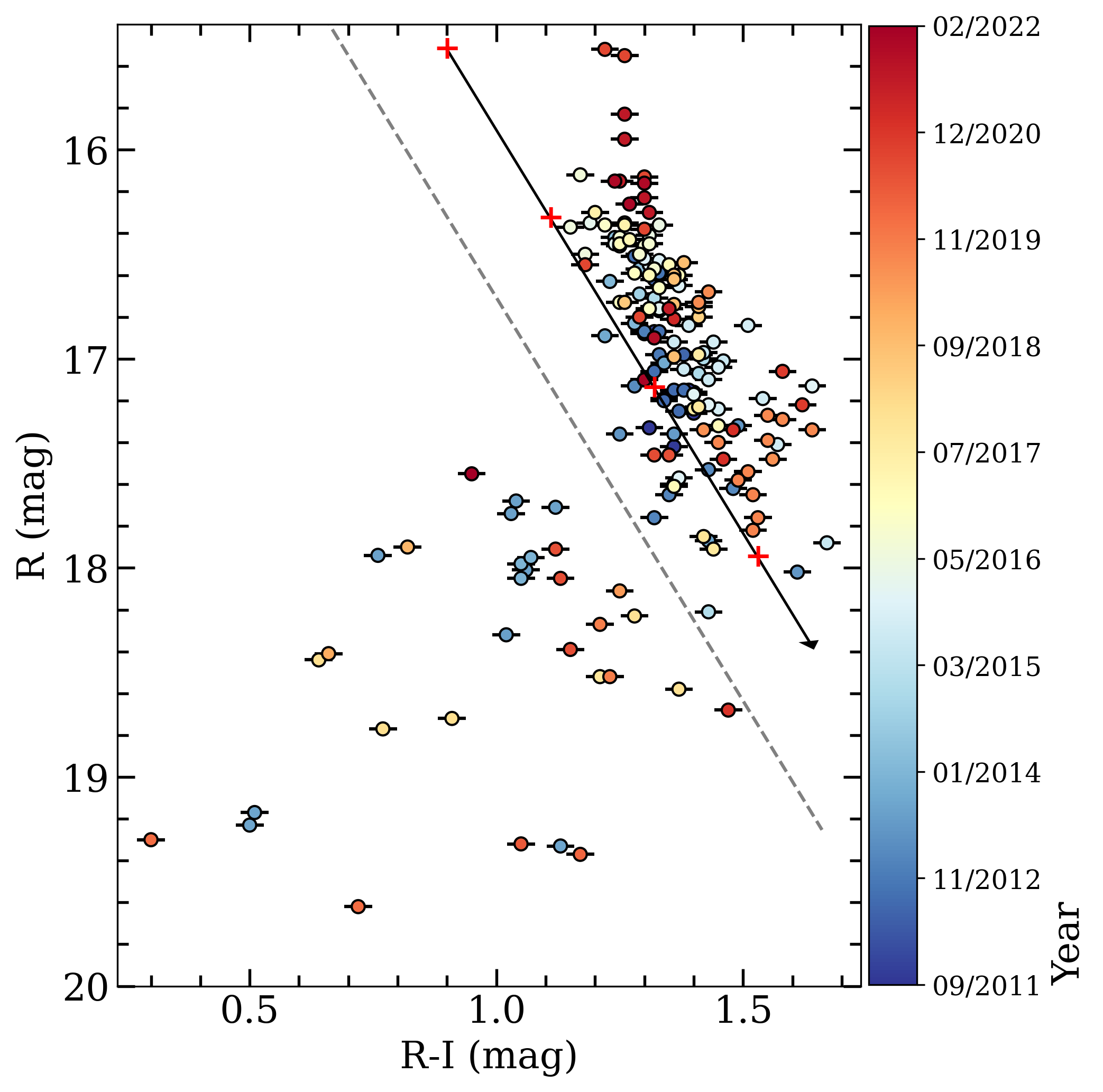}
    \includegraphics[width=0.3\textwidth]{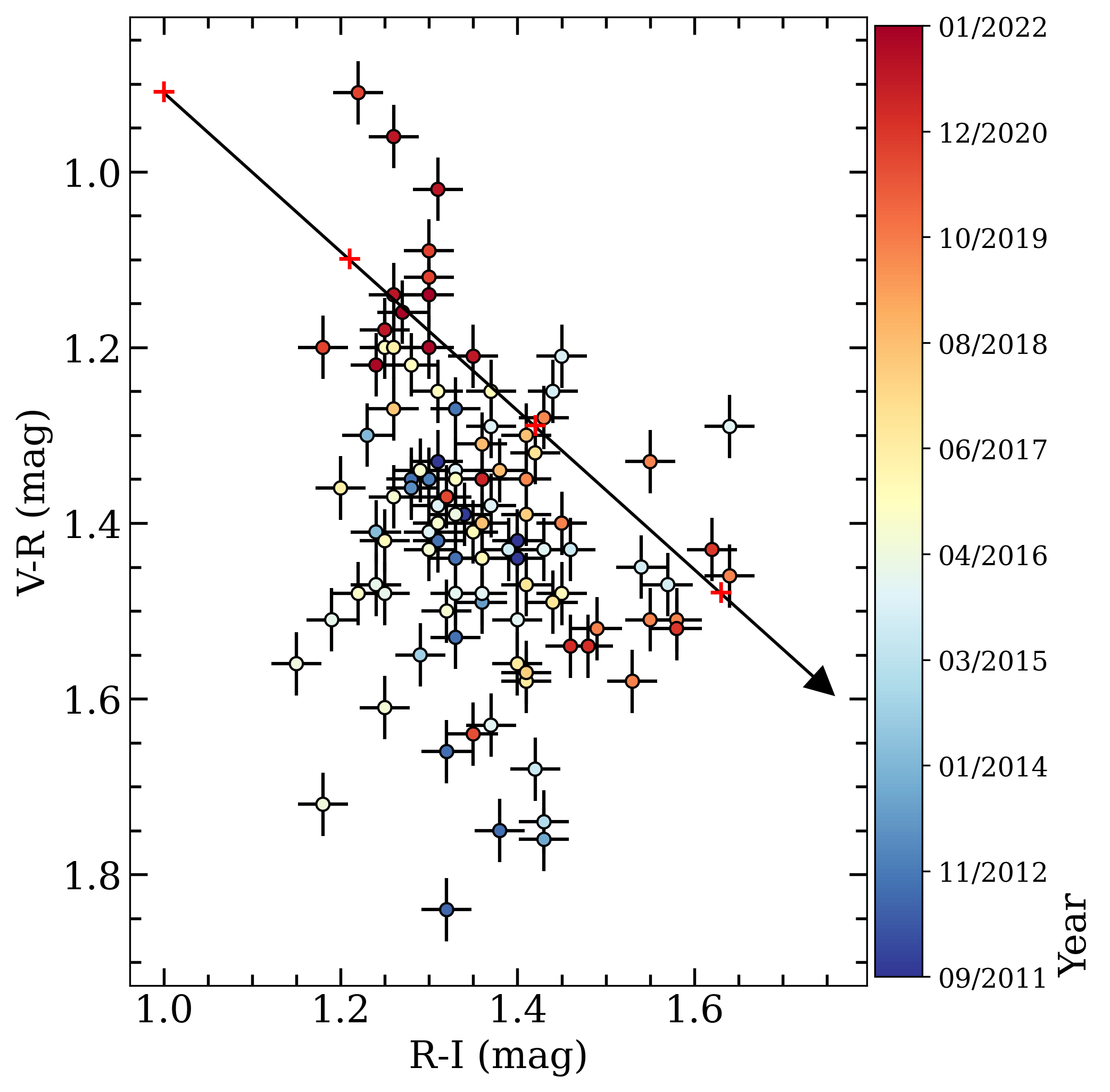}
    \caption{($V$ vs. $V-I$) and ($R$ vs. $R-I$) CMD (left and middle panel, respectively), using photometric data from \cite{Mutafov_2022RAA....22l5014M}. The gray dashed lines indicate the cutoff used to exclude data points lying below these boundaries, which are assumed to be affected by the \enquote{blueing effect}. The right panel shows ($V-R$ vs. $R-I$) CC diagram using data points which are above the cutoff. The black arrow shows $A_V$ vector with $\Delta A_V=1$ mag  marked as red \enquote{+} sign.}
    \label{fig:av_clc}
\end{figure*}

\renewcommand\thetable{A\arabic{table}}
\setcounter{table}{0}

\begin{table*}
    \centering
    \caption{Line fluxes measured from HFOSC spectra. Observation dates are formatted as 'YYYYMMDD'. OI$_{1}$ and OI$_{2}$ denote the [O \textsc{i}] forbidden lines at 6300 and 6363 \r{A}, respectively. CaII$_1$, CaII$_2$, and CaII$_3$ correspond to the Ca \textsc{ii} IR triplet lines at 8498, 8542, and 8662 \r{A}. SII$_{1}$ and SII$_{2}$ indicate the [S \textsc{ii}] forbidden lines at 6717 and 6731 \r{A}. All flux values are given in units of ($\times$10$^{-15}$ erg s$^{-1}$ cm$^{-2}$).}
    \tiny
    \begin{tabular}{ccccccccccccccccccc}
        \hline
        \hline
         Date & H${\beta}$ & eH${\beta}$ & OI$_{1}$ & eOI$_{1}$ & OI$_{2}$ & eOI$_{2}$ & H${\alpha}$ & eH${\alpha}$ & SII$_{1}$ & eSII$_{1}$ & SII$_{2}$ & eSII$_{2}$ & CaII$_1$ & eCaII$_1$  & CaII$_2$ & eCaII$_2$  & CaII$_3$ & eCaII$_3$  \\
         
  \hline
    20150818 &     -- &  -- &  21.2 &  1.0 &   7.2 &    1.2 & 306.2 & 11.8 &    2.0 &     0.4 &    4.1 &     0.5 &    18.5 &      0.3 &    21.7 &      0.3 &    19.3 &      0.4 \\
20150927 &   108.4 & 17.7 &  47.8 &  2.3 &  12.9 &    2.1 & 375.5 &  4.6 &    3.8 &     0.6 &    6.0 &     0.8 &     -- &      -- &     -- &      -- &     -- &      -- \\
20151014 &     -- &  -- &  31.3 &  1.2 &   9.0 &    0.8 & 422.4 &  6.2 &    2.7 &     0.4 &    4.6 &     0.5 &    11.7 &      0.4 &    13.7 &      0.3 &    12.4 &      0.3 \\
20151110 &     -- &  -- &  33.3 &  1.1 &  10.6 &    0.8 & 570.1 &  7.6 &    4.0 &     0.5 &    5.4 &     0.6 &    18.3 &      0.5 &    21.3 &      0.3 &    18.7 &      0.5 \\
20160120 &     -- &  -- &  28.6 &  1.8 &   7.8 &    0.9 & 491.4 & 13.3 &    1.6 &     0.4 &    3.8 &     0.5 &    18.3 &      0.7 &    20.9 &      0.6 &    18.9 &      0.7 \\
20160203 &     -- &  -- &  27.5 &  1.3 &   9.1 &    1.0 & 562.4 &  5.1 &    2.4 &     0.4 &    3.4 &     0.5 &    30.1 &      0.6 &    33.8 &      0.6 &    28.8 &      0.5 \\
20161006 &    96.7 &  4.3 &  24.1 &  1.6 &   6.8 &    0.9 & 560.1 &  6.9 &    -- &     -- &    -- &     -- &     -- &      -- &     -- &      -- &     -- &      -- \\
20161208 &    47.8 & 17.2 &  37.0 &  3.5 &  11.1 &    2.6 & 371.4 &  9.1 &    -- &     -- &    -- &     -- &    35.5 &      0.5 &    38.9 &      0.5 &    36.0 &      0.7 \\
20161210 &    39.4 & 12.2 &  14.6 &  2.4 &   3.9 &    1.8 & 227.0 &  4.2 &    -- &     -- &    -- &     -- &     -- &      -- &     -- &      -- &     -- &      -- \\
20170102 &   100.4 &  4.6 &  27.6 &  1.8 &  10.5 &    1.2 & 434.5 & 10.9 &    2.3 &     0.4 &    4.6 &     0.5 &    23.4 &      0.8 &    27.4 &      0.4 &    24.2 &      0.5 \\
20170918 &    68.2 & 13.3 &  32.2 &  2.0 &   9.7 &    1.3 & 127.8 &  4.2 &    3.5 &     0.4 &    5.4 &     0.5 &     2.4 &      0.2 &     2.5 &      0.1 &     2.6 &      0.2 \\
20171024 &    90.5 &  9.9 &  71.9 &  3.0 &  23.1 &    2.4 & 343.9 & 10.2 &    7.9 &     0.6 &   12.9 &     0.8 &    12.5 &      0.5 &    11.4 &      0.5 &    11.6 &      0.8 \\
20171126 &    19.7 &  3.0 &  37.5 &  1.3 &  12.1 &    1.0 & 221.8 &  5.3 &    4.4 &     0.5 &    7.1 &     0.5 &    17.6 &      0.6 &    18.0 &      1.7 &    18.7 &      0.8 \\
20180218 &    73.9 &  7.6 & 119.5 &  4.6 &  38.7 &    2.8 & 376.3 &  8.0 &   14.4 &     1.4 &   21.3 &     1.6 &     7.4 &      0.9 &     7.5 &      0.5 &     7.5 &      0.5 \\
20181121 &   248.7 & 29.9 &  19.2 &  1.5 &   5.8 &    1.6 & 128.7 &  3.0 &    -- &     -- &    -- &     -- &     4.3 &      0.2 &     3.8 &      0.4 &     4.0 &      0.2 \\
20181212 &    57.7 & 10.8 &  10.5 &  1.1 &   2.5 &    0.7 &  63.9 &  2.0 &    -- &     -- &    -- &     -- &     2.5 &      0.1 &     2.5 &      0.1 &     2.2 &      0.1 \\
20181225 &    28.0 &  3.2 &   5.3 &  0.6 &   1.7 &    0.5 &  67.4 &  1.3 &    -- &     -- &    -- &     -- &     3.3 &      0.1 &     3.3 &      0.2 &     2.9 &      0.2 \\
20190927 &     -- &  -- &  13.9 &  1.8 &   7.9 &    2.4 &  80.3 &  2.9 &    -- &     -- &    -- &     -- &     8.4 &      0.5 &     9.1 &      0.6 &     9.3 &      0.8 \\
20191008 &     6.1 &  2.9 &  17.5 &  1.9 &   2.9 &    1.2 & 120.4 &  4.1 &    -- &     -- &    -- &     -- &    12.8 &      0.6 &    14.6 &      0.6 &    12.7 &      0.7 \\
20191028 &    19.4 &  6.8 &  45.9 &  2.7 &  11.7 &    3.1 & 180.7 &  7.8 &    -- &     -- &    -- &     -- &     -- &      -- &     -- &      -- &     -- &      -- \\
20200914 &    49.9 &  4.3 &  16.6 &  1.2 &   7.5 &    1.6 & 350.3 &  4.8 &    2.6 &     0.7 &    3.2 &     0.8 &    69.1 &      0.8 &    78.4 &      1.0 &    67.8 &      0.7 \\
20210306 &    33.2 &  2.0 &  11.6 &  1.1 &   4.3 &    0.9 & 125.4 &  2.9 &    2.1 &     0.4 &    3.2 &     0.4 &    20.7 &      0.4 &    23.3 &      0.3 &    20.2 &      0.4 \\
20220207 &     -- &  -- &  49.8 &  3.6 &  15.7 &    1.7 & 583.5 &  8.7 &    6.7 &     0.6 &   11.0 &     0.8 &    34.7 &      1.2 &    38.1 &      1.8 &    31.5 &      1.0 \\
 \hline
    \end{tabular}
    \label{tab:ln_strngth_HFOSC}
\end{table*}
\normalsize

\begin{table*}
   \centering
    \caption{EWs calculated from HFOSC spectra. Observation dates follow the 'YYYYMMDD' format. OI is the oxygen line at 6300 \r{A}. CaII$_1$, CaII$_2$, and CaII$_3$ denote Ca II IR triplet lines at 8498, 8542, and 8662 \r{A}, respectively. All EWs listed in the table are in \r{A}.}
    \begin{tabular}{rrrrrrrrrrrrr}
        \hline
        \hline
              Date &  H${\beta}$  &  eH${\beta}$ &   OI &   eOI &  H${\alpha}$ &  eH${\alpha}$ & CaII$_{1}$ &  eCaII$_{1}$ &  CaII$_{2}$ &  eCaII$_{2}$ &  CaII$_{3}$ &  eCaII$_{3}$ \\
\hline
 20150818 &    -- &   -- &  -21.9 &  1.1 & -249.8 &   6.4 &   -21.6 &      0.4 &   -24.8 &      0.2 &   -22.5 &      0.5 \\
 20150927 & -114.3 & 129.1 &  -39.5 &  2.0 & -275.3 &  11.4 &     -- &      -- &     -- &      -- &     -- &      -- \\
 20151014 &    -- &   -- &  -24.5 &  1.1 & -276.3 &   6.1 &   -14.4 &      0.4 &   -15.6 &      0.4 &   -14.1 &      0.2 \\
 20151110 &    -- &   -- &  -22.6 &  1.6 & -317.1 &   6.4 &   -18.6 &      0.7 &   -22.2 &      0.5 &   -19.8 &      0.8 \\
 20160120 &    -- &   -- &  -25.0 &  2.0 & -330.6 &   9.1 &   -21.2 &      0.7 &   -23.9 &      0.5 &   -22.7 &      0.9 \\
 20160203 &    -- &   -- &  -21.5 &  1.1 & -334.4 &   6.9 &   -31.8 &      0.5 &   -35.9 &      0.6 &   -30.4 &      0.6 \\
 20161006 &  -59.9 &   5.8 &  -16.4 &  0.7 & -282.0 &   9.0 &     -- &      -- &     -- &      -- &     -- &      -- \\
 20161208 &  -15.1 &   5.8 &  -30.7 &  3.8 & -247.1 &   8.8 &   -42.7 &      1.2 &   -47.4 &      1.1 &   -48.0 &      1.3 \\
 20161210 &  -19.0 &   4.5 &  -12.1 &  1.4 & -160.5 &   5.2 &     -- &      -- &     -- &      -- &     -- &      -- \\
 20170102 &  -56.2 &   4.6 &  -18.6 &  0.9 & -242.4 &   5.2 &   -25.3 &      0.6 &   -29.7 &      0.5 &   -25.2 &      0.5 \\
 20170918 & -109.6 &  40.8 & -105.8 & 13.4 & -343.5 &  32.2 &   -18.2 &      2.1 &   -20.0 &      1.5 &   -18.9 &      2.4 \\
 20171024 & -160.6 &  57.3 & -188.3 & 15.7 & -514.4 &  46.5 &   -31.9 &      3.0 &   -31.9 &      2.1 &   -42.0 &      6.7 \\
 20171126 &  -98.9 &  29.1 & -171.4 & 23.0 & -586.5 &  55.6 &   -71.7 &      3.8 &   -87.2 &      4.7 &   -89.0 &      5.4 \\
 20180218 &  -81.2 &  20.7 & -261.6 & 35.5 & -511.7 &  71.3 &   -24.5 &      3.1 &   -27.6 &      2.6 &   -27.2 &      2.5 \\
 20181121 & -315.8 & 123.2 &  -53.7 &  6.8 & -306.5 &  21.5 &   -32.9 &      1.4 &   -33.6 &      2.0 &   -34.2 &      3.1 \\
 20181212 &  -35.8 &   9.7 &  -25.2 &  3.1 & -165.2 &   7.7 &   -20.8 &      1.3 &   -21.3 &      1.0 &   -22.6 &      1.4 \\
 20181225 &  -22.3 &   3.8 &  -14.0 &  1.3 & -207.8 &   6.0 &   -30.4 &      1.1 &   -31.9 &      2.0 &   -29.3 &      2.2 \\
 20190927 &    -- &   -- &  -29.9 &  3.8 & -205.2 &  13.3 &   -35.5 &      2.8 &   -45.9 &      2.3 &   -45.5 &      2.6 \\
 20191008 &   -9.2 &   4.9 &  -88.8 & 24.3 & -472.5 &  61.8 &   -55.6 &      4.0 &   -61.4 &      5.2 &   -66.3 &      7.9 \\
 20191028 &  -51.1 &  27.4 & -102.0 & 23.6 & -440.3 & 107.6 &     -- &      -- &     -- &      -- &     -- &      -- \\
 20200914 &  -19.0 &   2.7 &   -6.5 &  0.5 & -161.5 &   4.6 &   -44.4 &      2.1 &   -57.4 &      1.7 &   -51.8 &      1.9 \\
 20210306 &  -37.3 &   4.4 &  -12.0 &  0.7 & -124.9 &   3.2 &   -27.3 &      0.7 &   -33.0 &      0.6 &   -28.3 &      0.7 \\
 20220207 &    -- &   -- &  -71.6 & 11.7 & -479.2 &  28.0 &   -72.0 &      2.9 &   -87.7 &      3.3 &   -88.8 &      3.8 \\
 \hline
    \end{tabular}
    \label{tab:ew_HFOSC}
\end{table*}

\begin{table*}
   \centering
    \caption{Line fluxes calculated from TANSPEC spectra. Observation dates follow the 'YYYYMMDD' format. CaII$_1$, CaII$_2$, and CaII$_3$ denote Ca II IR triplet lines at 8498, 8542, and 8662 \r{A}, respectively. All values are expressed in units of ($\times$ 10$^{-15}$ erg s$^{-1}$ cm$^{-2}$).}
    \tiny
    \begin{tabular}{ccccccccccccccccccc}
        \hline
        \hline
         Date &  H${\alpha}$ & eH${\alpha}$ &  CaII$_{1}$ &  eCaII$_{1}$ &  CaII$_{2}$ &  eCaII$_{2}$ &  CaII$_{3}$ &  eCaII$_{3}$ & Pa9 & ePa9 & Pa7 & ePa7 & Pa6 & ePa6 & Pa5 & ePa5 &  Br$_{\gamma}$ &   eBr$_{\gamma}$ \\
  \hline
 20221102 &  245.2 & 11.4 &    28.7 &      2.2 &    27.1 &      1.6 &    23.5 &      1.8 &  3.9 &   0.8 &  4.1 &   1.1 &  5.8 &   0.7 &  7.3 &   0.3 &   1.2 &    0.1 \\
20240114 &  372.5 & 22.2 &    47.9 &      5.7 &    81.1 &     14.3 &    53.5 &      7.3 &  9.9 &   1.9 &  9.2 &   1.5 & 10.8 &   0.9 & 12.1 &   0.4 &   1.8 &    0.1 \\
20240121 &  440.7 & 15.2 &    58.9 &      1.9 &    71.6 &      2.8 &    55.7 &      3.5 &  5.0 &   1.0 &  7.0 &   0.8 & 11.1 &   0.7 & 14.8 &   0.4 &   1.9 &    0.1 \\
20240122 &  382.9 & 29.1 &    53.7 &      5.8 &    70.5 &      3.4 &    60.7 &      5.2 &  8.7 &   1.6 &  5.7 &   1.4 & 10.2 &   1.5 & 12.4 &   1.0 &   2.0 &    0.1 \\
20240123 &  267.9 & 12.1 &    39.6 &      1.7 &    48.7 &      2.5 &    41.7 &      2.4 &  4.9 &   0.9 &  6.1 &   0.8 &  9.2 &   0.5 & 12.9 &   0.3 &   2.3 &    0.1 \\
20240124 &  520.4 & 15.7 &    79.9 &      3.1 &   100.7 &      2.8 &    85.4 &      5.3 &  9.6 &   1.4 & 12.6 &   1.2 & 14.9 &   0.8 & 18.2 &   0.6 &   2.5 &    0.1 \\
20240209 &  371.4 & 18.6 &    74.2 &      2.9 &    85.9 &      3.1 &    83.6 &      3.0 &  6.7 &   0.8 & 12.7 &   1.8 & 14.0 &   0.9 & 14.2 &   0.4 &   2.3 &    0.1 \\
 \hline
    \end{tabular}
    \label{tab:ln_strngth_TANSPEC}
\end{table*}
\normalsize

\begin{table*}
   \centering
    \caption{EWs calculated from TANSPEC spectra. Observation dates follow the 'YYYYMMDD' format. CaII$_1$, CaII$_2$, and CaII$_3$ denote Ca II IR triplet lines at 8498, 8542, and 8662 \r{A}, respectively. All EWs listed in the table are in \r{A}.}
    \tiny
    \begin{tabular}{ccccccccccccccccccc}
        \hline
        \hline
              Date &       H${\alpha}$ &      eH${\alpha}$ &  CaII$_{1}$ &  eCaII$_{1}$ &  CaII$_{2}$ &  eCaII$_{2}$ &  CaII$_{3}$ &  eCaII$_{3}$ &    Pa9 &  ePa9 &    Pa7 &  ePa7 &    Pa6 &  ePa6 &    Pa5 &  ePa5 &  Br$_{\gamma}$ &  eB$_{\gamma}$\\
\hline
 20221102 &  -137.2 &    24.2 &   -41.4 &      6.3 &   -42.3 &      6.5 &   -36.2 &      6.7 &  -5.7 &   1.0 &  -7.6 &   1.6 &  -9.7 &   1.1 & -15.0 &   0.4 &  -4.2 &    0.6 \\
 20240114 &   -91.7 &    17.4 &   -44.3 &      6.1 &   -54.0 &     21.2 &   -34.1 &      6.1 &  -9.0 &   2.2 & -10.4 &   2.0 & -13.7 &   1.1 & -16.9 &   0.6 &  -4.8 &    0.4 \\
 20240121 &  -113.2 &    14.7 &   -33.7 &      1.8 &   -53.3 &      4.6 &   -33.1 &      2.5 &  -3.7 &   1.0 &  -8.1 &   1.2 & -12.8 &   1.0 & -18.4 &   0.5 &  -4.6 &    0.3 \\
 20240122 & -- & -- &   -37.8 &      7.9 &   -74.7 &     14.0 &   -63.2 &     16.5 & -11.9 &   2.6 &  -5.2 &   1.2 & -12.8 &   2.2 & -15.9 &   1.8 &  -4.6 &    0.3 \\
 20240123 &  -146.6 &    22.7 &   -31.9 &      2.6 &   -48.2 &      3.8 &   -34.6 &      2.9 &  -4.8 &   0.6 &  -8.6 &   1.4 & -12.6 &   0.8 & -19.1 &   0.6 &  -5.2 &    0.3 \\
 20240124 &  -146.5 &    13.8 &   -39.1 &      3.1 &   -56.5 &      3.6 &   -50.7 &      2.8 &  -6.4 &   0.9 & -13.2 &   1.5 & -14.2 &   0.9 & -19.2 &   0.7 &  -5.4 &    0.2 \\
 20240209 &  -121.3 &    18.5 &   -39.2 &      2.6 &   -47.5 &      2.8 &   -45.0 &      3.4 &  -4.0 &   0.5 & -12.5 &   1.7 & -12.1 &   0.5 & -14.7 &   0.8 &  -5.2 &    0.3 \\
 \hline
    \end{tabular}
    \label{tab:ew_TANSPEC}
\end{table*}
\normalsize

\begin{table*}
   \centering
    \caption{Accretion rates calculated from HFOSC spectra. Observation dates follow the 'YYYYMMDD' format. CaII$_1$, CaII$_2$, and CaII$_3$ denote Ca II IR triplet lines at 8498, 8542, and 8662 \r{A}, respectively. All values are expressed in units of ($\times$ 10$^{-8}$ $M_\odot$yr$^{-1}$).}
    \begin{tabular}{rrrrrr}
        \hline
        \hline
         Date & H${\beta}$ & H${\alpha}$ & CaII$_1$ & CaII$_2$ & CaII$_3$ \\
  \hline
 20150818 &   -- & 30.8 &    20.3 &    17.9 &    14.7 \\
 20150927 &  64.8 & 38.7 &     -- &     -- &     -- \\
 20151014 &   -- & 44.2 &    12.9 &    11.4 &     9.8 \\
 20151110 &   -- & 62.1 &    20.0 &    17.5 &    14.3 \\
 20160120 &   -- & 52.5 &    20.0 &    17.3 &    14.5 \\
 20160203 &   -- & 61.1 &    32.9 &    27.5 &    21.4 \\
 20161006 &  56.9 & 60.8 &     -- &     -- &     -- \\
 20161208 &  25.5 & 38.2 &    38.7 &    31.5 &    26.3 \\
 20161210 &  20.5 & 21.9 &     -- &     -- &     -- \\
 20170102 &  59.4 & 45.7 &    25.6 &    22.4 &    18.2 \\
 20170918 &  38.2 & 11.5 &     2.7 &     2.2 &     2.3 \\
 20171024 &  52.8 & 35.1 &    13.8 &     9.5 &     9.2 \\
 20171126 &   9.3 & 21.4 &    19.4 &    15.0 &    14.3 \\
 20180218 &  41.9 & 38.8 &     8.2 &     6.3 &     6.1 \\
 20181121 & 167.1 & 11.5 &     4.8 &     3.3 &     3.4 \\
 20181212 &  31.6 &  5.2 &     2.7 &     2.2 &     1.9 \\
 20181225 &  13.9 &  5.6 &     3.7 &     2.9 &     2.5 \\
 20190927 &   -- &  6.8 &     9.3 &     7.7 &     7.4 \\
 20191008 &   2.5 & 10.7 &    14.0 &    12.2 &    10.0 \\
 20191028 &   9.1 & 16.9 &     -- &     -- &     -- \\
 20200914 &  26.7 & 35.8 &    74.8 &    62.2 &    47.4 \\
 20210306 &  16.8 & 11.2 &    22.7 &    19.2 &    15.4 \\
 20220207 &   -- & 63.7 &    37.8 &    30.9 &    23.3 \\
 \hline
    \end{tabular}
    \label{tab:accrtn_HFOSC}
\end{table*}

\begin{table*}
   \centering
    \caption{Accretion rates calculated from TANSPEC spectra. Observation dates follow the 'YYYYMMDD' format. CaII$_1$, CaII$_2$, and CaII$_3$ denote Ca II IR triplet lines at 8498, 8542, and 8662 \r{A}, respectively. All values are expressed in units of ($\times$ 10$^{-8}$ $M_\odot$yr$^{-1}$).}
    \begin{tabular}{rrrrrrrrrr}
        \hline
        \hline
         Date &  H${\alpha}$ &  CaII$_{1}$ &  CaII$_{2}$ &  CaII$_{3}$ & Pa9 & Pa7 & Pa6 & Pa5 &  Br$_{\gamma}$ \\
  \hline
 20221102 & 10.2 &    13.3 &    10.0 &     8.0 &  6.4 &  3.3 &  4.2 &  2.9 &   2.6 \\
 20240114 & 15.9 &    28.5 &    25.8 &    16.9 & 16.7 & 10.1 &  9.5 &  5.3 &   4.0 \\
 20240121 & 17.8 &    26.4 &    25.6 &    17.9 &  6.6 &  9.3 &  9.3 &  6.4 &   4.2 \\
 20240122 & 17.0 &    23.2 &    24.9 &    20.6 & 16.6 &  5.3 &  8.3 &  5.3 &   4.1 \\
 20240123 & 10.6 &    18.0 &    17.9 &    12.8 &  5.9 &  8.4 &  7.3 &  5.5 &   5.5 \\
 20240124 & 24.4 &    38.1 &    34.5 &    27.3 & 15.5 & 18.3 & 13.6 &  7.9 &   6.1 \\
 20240209 & 16.2 &    35.1 &    28.6 &    25.4 & 10.0 & 17.4 & 11.8 &  6.0 &   5.6 \\
 \hline
    \end{tabular}
    \label{tab:accrtn_TANSPEC}
\end{table*}

\end{document}